# Impact of directional walk on atom probe microanalysis

B.Gault[1*], F. Danoix[2,3], K. Hoummada[4], D. Mangelinck[4], H. Leitner[3]

[1] Australian Centre for Microscopy and Microanalysis, The University of Sydney, Australia

[2] Groupe de Physique des Matériaux, Université de Rouen - CNRS, Rouen, France

[3] Department of Physical Metallurgy and Materials Testing, Montanuniversität Leoben, Austria

[4] IM2NP, Université Paul Cézanne - CNRS, Marseille, France

Corresponding author: baptiste.gault@sydney.edu.au - Tel: + 612 9351 7548

**Abstract.** In the atom probe microanalysis of steels, inconsistencies in the measured compositions of solutes (C, N) have often been reported, as well as their appearance as molecular ions. Here we propose that these issues might arise from surface migration of solute atoms over the specimen surface. Surface migration of solutes is evidenced by field-ion microscopy observations, and its consequences on atom probe microanalysis are detailed for a wide range of solute (P, Si, Mn, B, C, N). It is proposed that directional walk driven by field gradients over the specimen surface and thermally activated is the prominent effect.

1. **Introduction**

Field-ion microscopy (FIM) enables imaging of the surface of a needle-shaped specimen, subjected to a high voltage, via the field-ionization of rare gas atoms in the vicinity of the surface [1]. Contrast either arises from topographical features or due to differences in the elemental nature of the surface atoms. As the electric field is progressively increased, FIM permits investigation of the structure in-depth via field-evaporation of the top-most layers. Field evaporation is a combination of electric-field-induced ionization and desorption of the atoms originally constituting the specimen surface. FIM images generally exhibit a series of concentric terraces, known as poles, forming facets surrounding major crystallographic directions protruding at the surface. The intersections between these facets forms a pattern of zone lines that further reveal crystallographic information. These crystallographic features form field evaporation of the specimen[2-3] and correspond to the adaptation of the specimen shape to the combination of the electric field and temperature [3]. Thanks to its excellent intrinsic spatial resolution [4], field-ion microscopy (FIM) has been extensively used to provide unprecedented insight into atomistic surface migration of single atoms or atomic clusters over atomically clean surfaces, and in particular into the process of directional walk whereby local electric field gradients drive migration of surface atoms [5-14].

Building on the imaging capabilities of the FIM, atom probe tomography (APT) enables three-dimensional imaging of materials with near-atomic resolution and high chemical sensitivity[15-17]. APT relies on time-controlled field evaporation of surface atoms[18]. High-voltage or laser pulses, superimposed to a DC electric field, trigger the process of field evaporation and allow elemental identification by time-of-flight mass spectrometry. Ions are subsequently accelerated away from the surface and are collected by a position-sensitive detector [19] which allows for three-dimensional reconstruction of the atomic positions[20]. Recent breakthroughs in the design of instruments via the implementation of the local electrode[21], large angle reflectrons [22-23] or shortened flight path instruments[24] have enabled significant increase in the field-of-view of atom probes similar to that of field ion microscopy (100 to 200 nm across) and of its impact on materials characterisation. The technique has been extensively used for microstructural investigation of structural[25] and functional materials [26], especially since the renaissance of the pulsed-laser APT [27-28].

If surface migration was evidenced in FIM, it was assumed that this artefact was not impacting atom-probe data, due to slightly different electric field conditions during APT, and hindered by a lack of clear evidence. Furthermore, the region in which APT microanalysis was performed used to be precisely chosen, via FIM, so as to avoid crystallographic features that are generally location highly affected by spatial aberration and so-called local magnification [29]. However, the increase in the field-of-view of APT has enabled clear observation of poles and zone lines within the cumulative detector hit maps, often referred to as desorption maps, and Yao *et al.* recently showed that in the analysis of a low-alloyed steel, C was prone to segregating to these features [30]. They assumed that this was due to surface migration of C during the analysis, driven and facilitated by the presence of electric field gradients over the specimen surface, as already observed by FIM [31-32]. They also proposed that the interstitial nature of C in steels might favour this behaviour. If surface migration was confirmed, this effect might severely impact atom probe tomography, as artificial variation in the composition of solute atoms would appear within the tomographic reconstruction resulting from surface diffusion and extending over tens or hundreds of nanometres, as will be shown later in this article. So not only spatial resolution, but also quantitativity of the technique may be affected.

It is therefore important to better understand this phenomenon in order to account for its potential consequences. In the present article, we investigate other solute atoms (N, P, Si, B, …) in different base metals (Fe, Ni). The effect of the evaporation mechanism, either high voltage or laser pulsed mode, is also investigated. Original data treatment methods to facilitate this research are also proposed. We show that suitable analysis conditions, under which the effects of surface diffusion are limited, can be found but are likely to be very dependent on each material.

2. **Experimental**

To ensure a thorough investigation of the phenomenon, several alloys were used in this study. In these materials, the different solute atoms under scrutiny were assumed to be in solid solution, and thus randomly distributed throughout the volume. The compositions of the alloys used are:

Alloy #1 : Fe-500atppmNb-250atppmC-250atppmN – As quenched

Alloy #2 : Fe-1.5at%Mn-1.2at%Si-0.05at%P - As quenched

Alloy #3 : Fe-500atppmNb-500atppmN-20atppmC – As quenched

Alloy #4 : Fe-40at%Al-400atppmB

Alloy #5 : low grade pure Ni (99% pure, purchased from Goodfellow).

Specimens for APT analysis were prepared from ~ 10x.4x.4 mm rods or wire with a diameter of ~.25 mm using the two stage method. A preliminary electrochemical polishing, using a solution of 25% perchloric acid in acetic acid, was performed, followed by a electrochemical fine-polishing stage using a solution of 2% perchloric acid in n-butoxyethanol under an optical microscope at room temperature.

FIM was performed on a Cameca EcOTAP, and images recorded on a CCD camera (EOT Marlin). Field ion micrographs were obtained using a mixture of Ne ($4.10^{-5}$ Torr) and He ($1.10^{-5}$ Torr) at 20K, and using the maximum tip-to-screen distance available, so as to maximize the magnification.

Atom-probe analyses were performed on the same instrument using high-voltage pulses, and on two different Cameca LEAP 3000 HR, equipped with a large-angle reflectron, and one Cameca LEAP 3000 X Si with a straight flight path. In addition to standard voltage-pulsing mode, each LEAP instrument had pulsed-laser capabilities: laser pulses ~ 10ps long at a wavelength of 532nm, focused at the apex with a beam diameter <10 μm. The beam position was set to maximize the detection rate, which is the average number of atoms detected per pulse (given in at/pulse). The instrument enables control of the temperature +/- 5K down to ~20K. During the analysis, the voltage is controlled so as to keep the detection rate constant. Unless specified, the detection rate was always set to be $5\ 10^{-3}$ at/pulse. Specifics of each experiment will be detailed when needed.

Field evaporation not only induces the departure of single atoms. A fraction of the ions are detected on so-called multiple events where more than a single ion, although resulting from a single pulse, reached the detector. These could be due to a process known as correlated field evaporation [33] or to in-flight dissociation of molecular ions [34]. Yao et al. showed that filtering the data based on their multiplicity improves the signal to noise ratio and detectability of low concentration elements [30]. This procedure will be used within this article.

3. **Evidence of surface migration**

   *3.1. Field Ion Microscopy*

In Figure 1, a sequence of field-ion micrographs during field evaporation of a specimen of alloy #1 is displayed. As the voltage is raised, the specimen is progressively field evaporated, as demonstrated by the shrinking and successive removal of the terraces on the main poles, and the internal structure of the material in depth is revealed. At least two precipitates are imaged throughout this evaporation sequence, as highlighted by yellow and green ellipses. The former corresponds to a carbide[35-36] while the latter is a Guinier-Preston type zone of Nb nitride[37].

In the second image of the sequence, a brightly imaging solute atom, highlighted by the red circle, appears on the surface. While the matrix atoms are progressively field evaporated, this atom is retained on the surface. It is readily apparent that this atom progressively migrates towards the central (011) pole, following the progression of the atomic terrace edge. This behaviour of isolated solute atoms was observed several times during the course of the experiment (images not shown here) and in other investigations on similar materials[38]. Interestingly, the bright spots belonging to neither of the two precipitates do not seem prone to surface migration prior to their departure. This indicates different behaviours of atoms depending on their local atomic neighbourhood. The elemental nature of the moving atom cannot be deduced from FIM, but is most likely to be either N or C, as discussed below.

### 3.2. Atom probe

FIM enables visualization of the atoms while they are on the surface whereas atom probe tomography provides the position of the field evaporation site of the atoms projected onto the position-sensitive detector, as well as the mass-to-charge ratio of the ion and hence its chemical nature. In the analysis of Fe-based alloys containing relatively low levels of solutes, maps showing the cumulative detector hit positions, commonly referred to as desorption maps, often exhibit a pattern that is characteristic of the crystallography and specific orientation of the specimen, similarly to field-ion micrographs. Examples of such maps are displayed in Figure 2, in which high density regions appear in red and low density in blue. Zone lines and poles are readily visible, and those corresponding to main crystallographic directions were identified. Contrasts in desorption maps are induced by changes in the local hit density on the detector. The atomic density is inversely proportional to the magnification, which depends on the local curvature and hence electric field [39]. These maps provide precise information about the faceting of the specimen surface and on the local strength of the electric field.

Faceting is a well-known process due to the equilibration of the specimen to the electric field and temperature conditions. It is mostly observed in pure or low-alloyed materials at low temperatures (i.e. < 50K). In the case of alloys with higher solute content, the presence of numerous solute atoms over the surface with different evaporation fields, or binding energies, prevents the specimen to adopt and equilibrium steady-state shape [40].

The distribution of detector positions of C and N ions from the atom probe analysis of a specimen of alloy #1 across a range of temperatures is shown in Figure 3. Only the C and N ions detected on multiple events are displayed. These detector positions are superimposed upon a map showing the density of iron ($Fe^{1+}$, $Fe^{2+}$, $Fe^{3+}$) ions detected as part of multiple hits. As discussed in ref.[33], this procedure also highlights regions of higher local electric field, where correlated field evaporation is more likely to take place. Conversely to Figure 2(a) that account for all the detected ions, the maps of Figure 3 only display those coming on multiple events.

In this specimen, C and N are expected to be homogeneously distributed in the ferritic solid solution. Whatever the temperature used to perform the experiment, an obvious crystallographic-dependent distribution is observed, where positions of both the N and C are clearly correlated to regions of higher electric field. Mapping the different carbon molecular ions ($C^{1+}$, $C^{2+}$, $C_3^{2+}$, $C_2^{1+}$) within this data set did not reveal any difference between locations where these species are found, as revealed by Figure 4. All carbon atoms, whatever their molecular form, follow the same trend. It is worth noting that about 75% of the ions in the corresponding ranges are part of multiple events, and the level of background part in the single hits is high as discussed in ref. [30]. When plotted together with the multiples, ions on single exhibit a similar trend, with a high concentration of hits around major crystallographic features, despite containing mostly background counts. Hetero-species clusters containing N, and to a lesser extent C, are also observed in steels, and positions of $NbN^{2+}$ and $NbC^{2+}$ ions are displayed in Figure 5, together with Nb ions. While the distribution of Nb ions appears to be random (yellow dots in Figure 5), hetero-cluster locations are clearly correlated with regions of high C and N densities, i.e. in regions exhibiting a locally higher electric field.

The presented results clearly show that directional walk is occurring at the specimen surface during atom probe analysis, and affects differently each atomic species.

4. **Influence of the experimental conditions**

Surface migration, and more specifically directional walk, is known to be a thermally assisted process driven by electric field gradients over the surface [32]. The behaviour of each atomic species is dictated by its polarisability as an adatom. The influence of the electric field gradient and temperature on the process was quantitatively investigated. Aseries of experiments was conducted on a single specimen of alloy #2 in laser pulsing mode across a range of laser powers (.4, .2, .1, .05 nJ) at 40K, and across a range of temperatures (20, 40, 80, 120 K) at .4nJ per pulse, and finally across the same range of temperatures but in HV pulsing mode with 15% pulse fraction. The evaporation rate was kept constant at $5\times10^{-3}$ ions/pulse in average. Each experiment contained ~ 1.5 million ions. To ensure that the specimen had reached its equilibrium shape after each change in experimental conditions, the ions from the beginning of each experiment were filtered out of any analysis until a constant pattern was observed in desorption map.

For each set of experimental conditions, the distance to the first nearest neighbour (1NN) was computed based on the detector coordinates for three types solute atoms (P, Si, Mn) respectively. It is noteworthy that all the ions were considered here, not only the ions coming on multiple events. Each of the solutes exhibited very different behaviours. Nearest neighbours methods have often been applied to three-dimensional reconstructed atom probe data for analyses ranging from cluster-finding algorithms [41], point-density measurements[41], matrix [42] or cluster [43] composition measurement and specific interaction between atomic species [44]. Here, 1NN are computed in two-dimensions, using the impact coordinates over the detector surface. This method will be referred to as $1NN^{Det}$. To better enable interpretation the $1NN^{Det}$ distributions are then compared with a random distribution. This distribution was obtained via generation of a corresponding dataset which maintained the positional detector coordinates of each atom in the original experimental dataset but randomly permuted the mass-to-charge ratios, and hence atomic identities, so as to mimic the data expected from a random solid solution [45]. It is important to note that these distributions are computed in the detector space, where, considering a magnification of .5 x $10^6$, 1 mm corresponds approximately to ~2nm within the reconstructed volume.

In Figure 6 the $1NN^{Det}$ distributions normalized by the random distribution are presented for the three solutes for all the different experiments. Each column corresponds to one solute, Mn, P and Si respectively, and each row corresponds to one set of experiments. The first set of experiments were carried out in pulsed laser mode with a constant laser power while the temperature was progressively decreased from 120 to 20K. Experiments of the second set were performed at a constant temperature of 40K, and with decreasing laser power from .4 to .05nJ. Finally, the third set of experiments was obtained in HV pulsing mode, with 15% pulse fraction, and the temperature was tuned from 120 to 20K.

In Figure 6, a value of 1 indicates the considered solute is randomly distributed, as observed for Mn, whatever the experimental analysis conditions. Conversely, values significantly larger than 1 at small distances indicate a pronounced tendency for short range clustering on the desorption map. This is effect is clearly apparent for silicon and phosphorous. These elements are not randomly distributed as expected, indicating that they are prone to surface migration before evaporation.

Silicon, together with Mn is one of two substitutional solute elements investigated in this study. Irregularities in the detection of this element have been previously reported [46]. If in HV pulsing mode no significant surface migration of Si was observed, in pulsed laser mode, a pronounced peak appears in the graph. This is indicative of a significant deviation from randomness, across the range of temperature, as the laser power is raised to 0.4 nJ/pulse. At 40K, the effect almost disappears when the laser power is reduced to 0.2 nJ/pulse and the distribution appears conform to random.

The behaviour of P is similar to that of Si, however, the effect is even more pronounced and is still observable even with the lowest 0.05 nJ/pulse laser power at 40K. This suggests that P is more sensitive to surface migration than Si. It is now recognized that laser irradiation of APT tips results in a temperature increase, proportional to the laser power [47] and hence the dependence of surface migration with the laser power indicates that this phenomenon is, at least partly, thermally activated.

The results indicate that both substitutional (here Si) and interstitial (here P) species can be susceptible to surface diffusion. Nevertheless, interstitial species investigated in this study are clearly more prone to be affected, as shown by the normalized $1NN^{Det}$ values measured for P in alloy #2 (Fig.6), and N and C in alloy #1 (Fig.3). In addition, similarly to what was observed for Nb in alloy #1 in a previous section, Mn does not seem to be affected. The reason why some substitutional species are selectively affected by surface diffusion will be further discussed later in this paper.

As proposed by Haydock and Kingham, ions formed by field evaporation can be subsequently ionized by the electric field in the vicinity of the surface[48-49]. This translates into a higher probability of finding higher charge-states in regions of higher local electric field that can occur across the surface of the tip. Using a curve computed from Kingham's data and model[49], a map of the electric field strength across the tip surface can be estimated via the charge-state ratios of Fe at each point in the desorption map. This enables estimation of the amplitude of possible electric field gradients existing over the specimen surface. Examples of such maps are shown in Figure 7 in the cacse of alloy #2. The obvious changes in the electric field distribution as the temperature and the laser power are raised are indicative of significant change in the shape of the specimen. Even though poles and zone lines can be distinguished, the map obtained in HV pulsing mode is mostly homogeneous as shown in the bottom right of Fig. 7 (red border).

In parallel, the charge-state ratio measured for the series of runs as a function of the temperature in HV pulsing mode was used to establish a calibration of the effective temperature at which field evaporation takes place $T_{eff}$, as proposed by Kellogg[50]. The charge state ratio of Fe was determined based on all the ions within data set disregarding their spatial distribution and was plotted as a function of the base temperature. A polynomial was then fitted to experimental curve. The fitted curve was then used to estimate the temperature at which the field evaporation would take place for each of the experimental conditions. It is known that this method only yields rough estimates of the temperature, especially in the case of alloys [51], even more since some of the values of the temperature deduced here lie out of the initial range of base temperature investigated. Notwithstanding, it is still one of the only ways to assess the experimental conditions at the specimen surface. The amplitude of variation in the electric field ΔF across the detector is plotted as a function of $T_{eff}$ in Figure 8. Note that aberrant values of the field due to effect at the boundary of the map were not considered in estimating the gradient.

Changes in ΔF of a factor 2 are observed depending on the analysis conditions. Interestingly, this graph shows some consistency in the values of the field gradient observed in laser pulsing (diamonds) and HV pulsing at high temperature (circles), which probably translate in similar specimen shape. These changes in the field gradient with the analysis conditions are most likely to explain the trend observed in Figure 6, as conditions that minimise both the field gradient and effective temperatures yield homogeneous distributions of P, Si and Mn altogether.

5. Influence on the composition measurement

The most significant impact of directional walk on atom probe data analysis is the resulting differences in local composition it can cause. Atoms migrating towards crystallographic features induce artificial local surface concentration variations, with peaks of high solute concentration surrounded by depleted regions.

This effect was most likely present, even if not recognized at the time, within atom probe analyses using smaller fields of view, like in the original 1D atom probe, or even in the first generation of 3D atom probes. In particular, previous studies of alloy # 3 via small angle atom probe tomography (EcOTAP), exhibited variations of C and N from one analysis to the other, conducted on different specimens. The composition in these data sets was measured and the ratios in the minimal and maximal compositions of C and N were respectively $C_{max}/C_{min}=2.1$ and $N_{max}/N_{min},=3.5$, while the corresponding Nb ratio was 1.3. These variations were previously attributed to local heterogeneities within the material, but, with the present results, and thanks to the much wider field of view, another obvious explanation can be proposed. As the analysis surfaces were one to two orders of magnitude smaller, analyses were probably conducted randomly on either high field or low field regions, resulting in large discrepancies in measured local solute contents. This is illustrated in Figure 9 which shows the distribution of N atoms in a wide field-of-view atom probe analysis of alloy #3. The circles on Fig 9 represent the typical size of surfaces analyzed by EcOTAP. Individual circles were positioned at selected location, corresponding to high (orange circle) and low (red circles) N level regions. Two other types of sites were also defined (green and blue regions). Local compositions within each type of zone are reported table 1. The max to min concentration values are respectively 1.5, 10 and 2 for Nb, N and C. These values show the same trend as was experimentally observed with EcOTAP, which supports the proposed explanation. The nitrogen concentration appears significantly higher around poles and zone lines. Also, no background correction was applied to the raw data, and as C appears in 6 different mass ranges, its composition is prone to overestimating. However, this in no way detracts from the conclusion that depending on the site where small angle atom probe analyses are performed large discrepancies in measured compositions can be encountered.

Similar results were obtained on Si and P in the series of analyses of alloy #2 discussed above. The relative peak values of the concentrations of Si and P are reported table 2. Note that the peak value is the maximum value measured in a 1x1 $mm^2$ square over the detector surface, i.e. about 2x2 $nm^2$ at the specimen surface. Ratios of more than 6 are observed for the first set of experiments, the higher values measured for the higher temperatures and laser powers. These results confirms that laser pulsing exacerbates the surface concentration variations.

The trend of a decrease in Si composition as the temperature increases in HV pulsing mode is opposite to the trend observed by Miller and Smith in Si-containing steels and that was attributed to DC field evaporation of Fe, inducing a preferential loss of Fe [46]. However, considering the high crystallographic dependence of surface migration related processes and that their results were only obtained around the (111) pole, the difference in trend might be due to difference in the surface migration behaviour of Si on Fe.

Once variations are identified, the question that arises is whether any selected analysis region could nevertheless provide sensible results. When compared to the nominal composition as a reference, it appears that D type zones are less affected. One could therefore think, from a wide angle dataset, selecting this type of regions could be an accurate way of measuring solute content. This result is only from one experiment, and such an experimental protocol has still to be proved efficient in other configurations (base material, experimental conditions, laser irradiation), but is clearly a promising way to overcome, at least partly, the deleterious effect of directional walk.

6. Discussion

Since the early work of Tsong and Kellogg [52] on directional walk of adsorbed atoms (adatoms), it is known that the energy barrier is species-specific. It is indeed proportional to the polarisability and hyperpolarisability of the adatom, for which values are different from those of the free-atom and are very poorly documented. This species specific was observed throughout our investigation, especially for high-evaporation field solute species like C and N. It is generally accepted that in the case where a solute atom requires a higher electric field to be field evaporated from the surface, it will be retained over the surface until the local electric field in its vicinity is sufficiently high to provoke its field evaporation. High-evaporation-field solutes will thus remain for a relatively longer time over the surface, enabling surface migration to take place.

Phosphorus and boron are two other high-evaporation field interstitial solutes. Further, if P was shown here to be severely prone to surface migration, B in alloy #4, did not exhibit the same behaviour, and only very shallow B concentration variations over the surface were observed. Alloy #4 also contains traces of P that were observed to migrate, which indicates that the relative surface migration is easier for P than for B. In addition to interstitial atoms, the case of Si in alloy#1 also demonstrates that substitutional solutes can also be prone to surface migration, even though Si migration was observed in high-temperature and high field gradient conditions. Nevertheless, not all substitutional atoms are subject to surface diffusion. In particular Mn and Nb have been investigated, and no such effect was observed. The lack of data on surface polarisability of these different elements on Fe prevents any clear explanation to be proposed and unambiguous conclusions to be drawn.

The behaviour observed mostly for substitutional atoms could be related to their individual atomic radius, which is much smaller than that of Fe. Surface migration might indeed be facilitated for smaller atoms, which can have more stable sites to migrate over the surface. Here is a list of atomic radius values taken from ref. [53]:
N (56 pm) < C (67 pm) < P (98 pm) < Si (111 pm) < Fe (156 pm) < Mn (161 pm) < Nb (198 pm)

The trends observed through the present study seem to be consistent with the atomic radii. N, C and P are the smallest atoms and the most prone to migration, followed by Si, Mn and Nb, these last two which have never observed migrating, being the two largest atoms in the present study. The difference in behaviour observed for different solutes or around different crystallographic features (poles, zone axes) could also be explained by the presence of specific sites that permit migration of certain solutes along specific paths and hence a strong influence of the crystallography on the surface migration process as already pointed out by several authors [9, 13-14, 54]. Boron constitutes an exception to this rule, but the investigation was performed in an ordered alloy with a different crystal structure which might impede B migration over the surface, notwithstanding the fact that chemical effects are likely to play a role.

Similarly to what was observed for metallic clusters migration on W surfaces[55], the migration and desorption of dimers can be easier than for single isolated atoms [56]. N and C are often observed as molecular species, forming both homo- and hetero- species cluster. This concept has been discussed for quite a long time, and was thought to underpin large discrepancies in the measurement of C composition and lowers the spatial resolution [57]. Thanks to the presence of a second natural isotope, it is possible to deconvolute the mass peaks corresponding to C molecular ions, and correct the measurement of atomic concentration [57]. Phosphorus being mono-isotopic, this procedure cannot be applied, but as no peak is observed at 31 and 62 Da, it is unlikely that $P_2$ molecular ions are formed and bias P concentration measurements. These molecular clusters are likely to be post-ionized in the vicinity of the specimen surface

[58], making them unstable and likely to dissociate. Interestingly, we did not observe a clear correlation between the different peaks of C (at 6,12,18,24 Da) and the different loci over the surface, which can be explained by the relatively limited differences between the electric field in these regions. Electric field induced dissociation of molecular ions results in a high proportion of these solutes arriving as part of multiple event on the detector, as observed in ref. [30], which might affect the accuracy of concentration measurement for these species.

The specific crystallography of the specimen seems to play a prominent role. Indeed, solute migration seems to be more pronounced around the (111) pole than the (011) for example. Also, solutes were often observed to align along specific directions, P was observed along the [011] [-111] direction. This crystallographic dependence was expected, as the surface diffusion coefficient and mechanism both strongly depend on the local arrangement of atoms over the surface[9]. The very same behaviour was observed in alloy #5, a low-grade pure Ni containing C and N. C and N were observed mostly concentrated at the (001) and (111) poles. Surface migration seemed to affect the data almost regardless of the experimental conditions, crystallographic structure (BCC for Fe and FCC for Ni) or the chemical nature of the substrate.

In the case of semiconductors, the very same problems are very likely to affect the data, especially since most APT analysis are performed in laser pulsing mode. Surface migration of various adatoms on Ge [59-60] and Si [61-62] was investigated by several techniques. The difference in the energy barrier for surface migration and desorption was measured to be in the same range (respectively 1.3-2.6 eV and 1.5-3.5 for example for Si[62]). This means that such processes are likely to similarly affect the atom probe microanalysis of some semiconductors, when a pole structure is observed within the desorption map [63].

Finally, due to the faceting, ΔF increases with the effective field evaporation temperature. This implies that the driving force and the thermal activation are simultaneously increased, greatly facilitating directional walk of solutes over the surface. It is worth noting though, that as the temperature is lowered, ΔF seems to reach a plateau and then increase at low temperature. This can be explained by changes in the specimen shape, with a more pronounced faceting at low temperature, inducing larger field gradients. As the energy barrier for field driven surface migration is strongly influenced by the electric field, as suggested by simulations [64-65] and experimental work[60], conditions might exist where directional walk will affect atom probe data even at very low temperature. This is counterintuitive but was observed in some of the data, as depicted in Figure 10. Again, a species-specific behaviour is expected, since the energy barrier for directional walk depends on the adatom polarisability and size.

7. **Conclusions**

In conclusion, surface diffusion processes, especially directional walk, was shown to occur for a wide range solute species. This phenomenon primarily affects interstitial elements, but has also been observed for substitutional elements such as Si in steel, impacting the microanalytical capabilities of APT. First, as solutes are migrating towards high-electric field regions over the specimen surface, the spatial resolving power of the technique is lowered. Secondly, solute often form homo- or hetero-species molecular ions to facilitate their departure from the surface, which can have a significant impact on the chemical accuracy. Other surface processes (i.e. surface reconstruction, kink-site detachment) were assumed to limit the spatial resolution in the case of pure metal [66-67]. They are also likely to take place at the surface of alloys such as those investigated here. This would mostly affect the position of matrix atoms and thus be less detrimental to the performance of APT. An in-depth resolution of ~100pm using only the Fe ions in three analyses of alloy #1 was measured around the (011) pole using the method introduced in ref.[68] based on the use of spatial distribution maps techniques[69]. Unfortunately, similar techniques can not be used to assess the exact resolution achievable for the solutes, as it is impossible to determine the original location occupied by the atom that have been subject to directional walk processes. Laser pulsing seems to make these surface migration effects more prominent, as they are thermally activated. They might also be promoted by the larger field gradients arising from enhanced faceting of the specimen at low temperature. For each alloy, experimental conditions where surface migration is either avoided or, at least, limited, should exist and be revealed by a systematic investigation. Further investigations are clearly needed to determine the best experimental conditions necessary for keeping the ultimate analytical performances of the instrument. On another point of view, careful quantification of this phenomenon might enable quantification of the surface polarisability of adatoms, data hardly obtainable with other techniques.

**Acknowledgements**

Baptiste Gault acknowledges the financial support of the European Commission via a FP7 Marie Curie IEF Action No. 237059. Thanks to Dr Michael Moody for proof reading the manuscript and fruitful discussions. The French National Research Agency (ANR is granted for financial support under contract CONTRA-PRECI # |0|6|-|B|L|A|N|-|0|2|0|5|. Ecole des Mines de Saint Etienne (France) is acknowledged for providing the FeAlB specimens.

**Figure caption**

Figure 1: Sequence of field ion micrographs during field evaporation of #1 alloy, aged 10 minutes at 650°C. The main poles were identified in the first image. Migration of a bright spot along a zone axis towards the (110) pole is evidenced by the red circles, while the yellow and green ellipses highlight the presence of precipitates within the micrographs. This bright spot most likely corresponds to either N or C.

Figure 2: desorption map for experiments at 20K and 20%PF in HV pulsing mode for specimens of alloy #1 and #3, respectively oriented along (110) in (a), and (111) in (b).

Figure 3: Distribution of C (green squares) and N (yellow stars) ions during the analysis of a specimen of alloy #1, across a range of temperatures in HV pulsing mode. Each map contains 1M ions with no precipitates. The distribution is superimposed to a two-dimensional density map of Fe ions detected on multiple hits that highlights regions of high electric field.

Figure 4: Distribution of carbon atomic and molecular ions during an analysis of an alloy #1 specimen at 40K around the (110) in (a) and of alloy #3 at 20K around the (111) pole in (b).

Figure 5: Same specimen as in Figure 3, experiment at 80K, distribution of NbN (red diamonds), NbC (blue triangles) molecular ions and Nb (yellow dots) ions.

Figure 6: Normalized $1NN^{Det}$ distribution of the detector impacts as a function of the distance for Mn, Si and P for the whole series of experiments in laser mode as a function of temperature and as of laser energy, and in HV pulsing mode as a function of temperature.

Figure 7: Maps of the electric field estimated from the charge state ratios for several analysis conditions in laser (blue and green border) and HV pulsing mode (red border).

Figure 8: Graph of ΔF as a function of the effective field evaporation temperature $T_{eff}$ for the series of experiments at constant laser power and varying temperature (square), varying laser power at constant temperature (diamond) and in HV pulsing mode at varying temperature (circle).

Figure 9: Top view of an analysis of alloy #3, where only the N atoms (blue dots) are represented. Zones mimicking the field-of-view of the EcOTAP are defined by circles of different colours. They were used to investigate variations of composition across the field-of-view reported in Table 1.

Figure 10: (a) Graphs similar to those of Figure 3 for the specimen showed in Figure 2(b), for two different temperatures. (b) Histogram of the distances to the centre of the (111) pole (equivalent to a species specific radial distribution function) for the different solutes at 20 and 70K.

Table 1: Nominal composition and measured composition in the different area defined in Figure 9.

Table 2: Summary of the different experiments treated in the article, including the experimental conditions, field gradient and local increase in solute concentration defined as the ratio between the maximum concentration measured in a bin on the detector and the average concentration.

**Table 1**

|  | Nb | N | C |
|---|---|---|---|
| Nominal composition (at.%) | 500 | 500 | 50 |
| A zone | 400 ± 50 | 2500 ± 200 | 180 ± 40 |
| B zones | 340 ± 50 | 260 ± 50 | 220 ± 40 |
| C zones | 440 ± 50 | 950 ± 150 | 300 ± 50 |
| D zones | 370 ± 50 | 550 ± 100 | 150 ± 35 |

**Table 2**

| Alloy # | Pulsing mode | Energy / PF | Base temperature (K) | $T_{eff}$ (K) | $\Delta F$ | $\Delta P_{max}$ | $\Delta Si_{max}$ |
|---|---|---|---|---|---|---|---|
| 2 | laser | .4 nJ | 120 | 304 | 1.25 | 5.04 | 1.98 |
| 2 | laser | .4 nJ | 80 | 169 | 0.85 | 5.55 | 2.15 |
| 2 | laser | .4 nJ | 40 | 153 | 0.75 | 5.53 | 2.14 |
| 2 | laser | .4 nJ | 20 | 156 | 0.80 | 5.37 | 1.88 |
| 2 | laser | .4 nJ | 40 | 168 | 0.81 | 6.17 | 2.11 |
| 2 | laser | .2 nJ | 40 | 115 | 0.60 | 5.53 | 1.20 |
| 2 | laser | .1 nJ | 40 | 77 | 0.65 | 3.86 | 0.99 |
| 2 | laser | .05 nJ | 40 | 63 | 0.61 | 1.31 | 1.22 |
| 2 | HV | 15 | 40 | 40 | 0.62 | 1.28 | 2.07 |
| 2 | HV | 15 | 80 | 80 | 0.61 | 1.29 | 1.05 |
| 2 | HV | 15 | 120 | 120 | 0.75 | 1.58 | 0.85 |
| 2 | HV | 15 | 20 | 20 | 0.71 | 2.26 | 3.01 |

| Alloy # | Pulsing mode | Energy / PF | Base temperature (K) | $T_{eff}$ (K) | $\Delta F$ | $\Delta C_{max}$ | $\Delta N_{max}$ |
|---|---|---|---|---|---|---|---|
| 1 | HV | 20 | 70 | 70 | 0.63 | 1.48 | 4.20 |
| 1 | HV | 20 | 20 | 20 | 0.67 | 2.09 | 6.39 |
| 3 | HV | 20 | 80 | 80 | 0.62 | 2.05 | 5.68 |
| 3 | HV | 20 | 40 | 40 | 0.64 | 1.88 | 2.20 |
| 3 | HV | 20 | 20 | 20 | 0.59 | 3.37 | 2.13 |

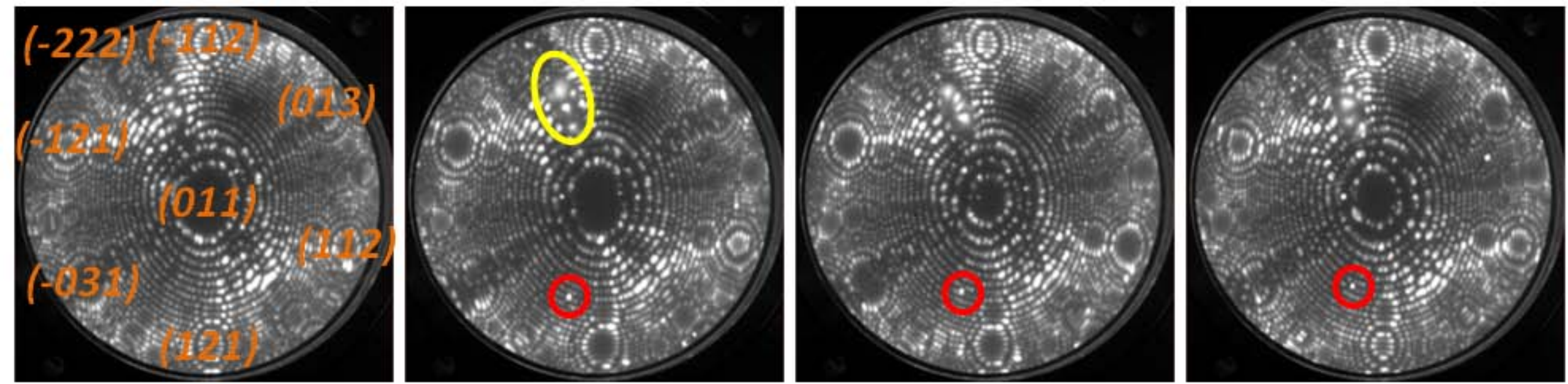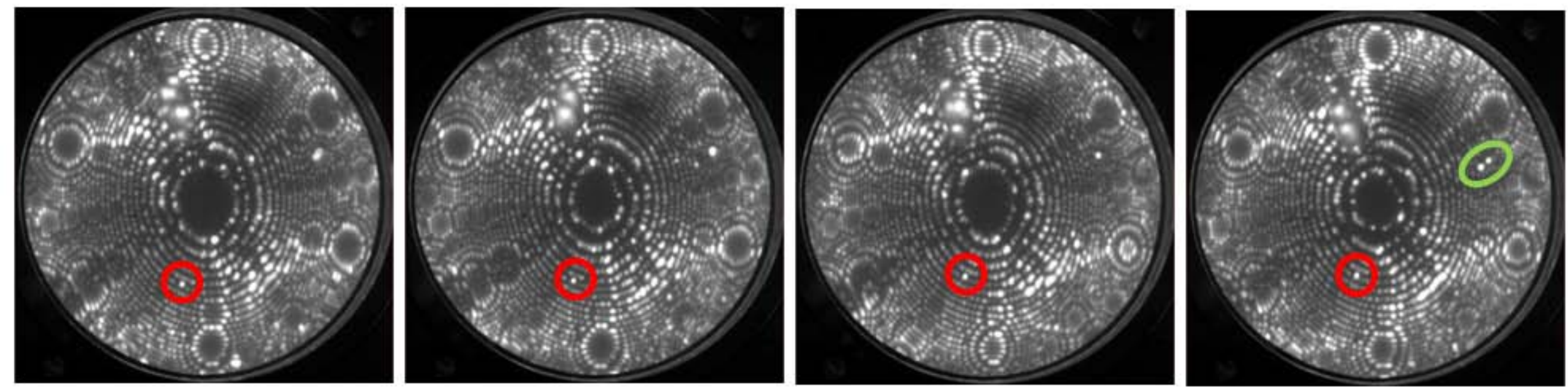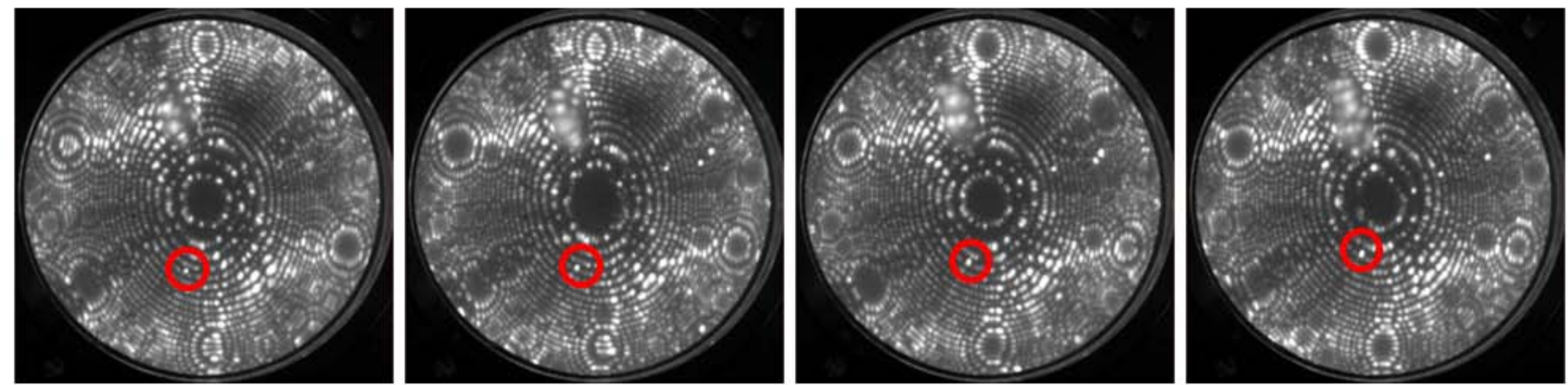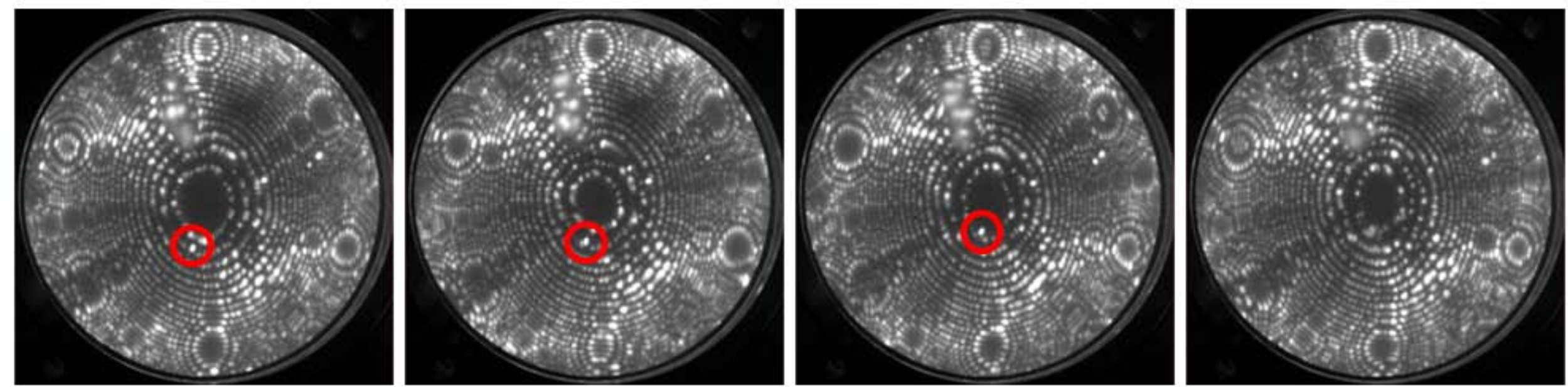

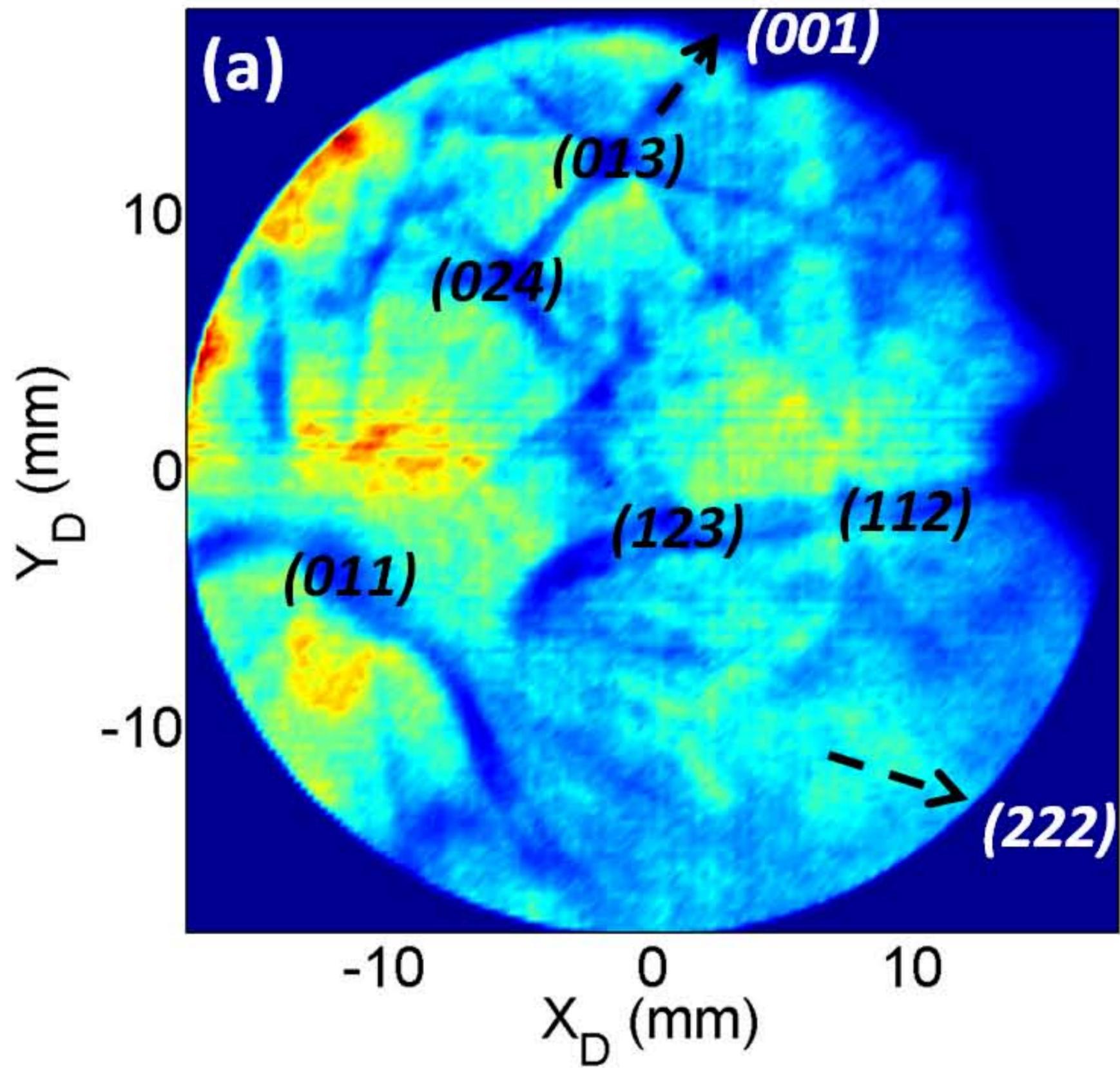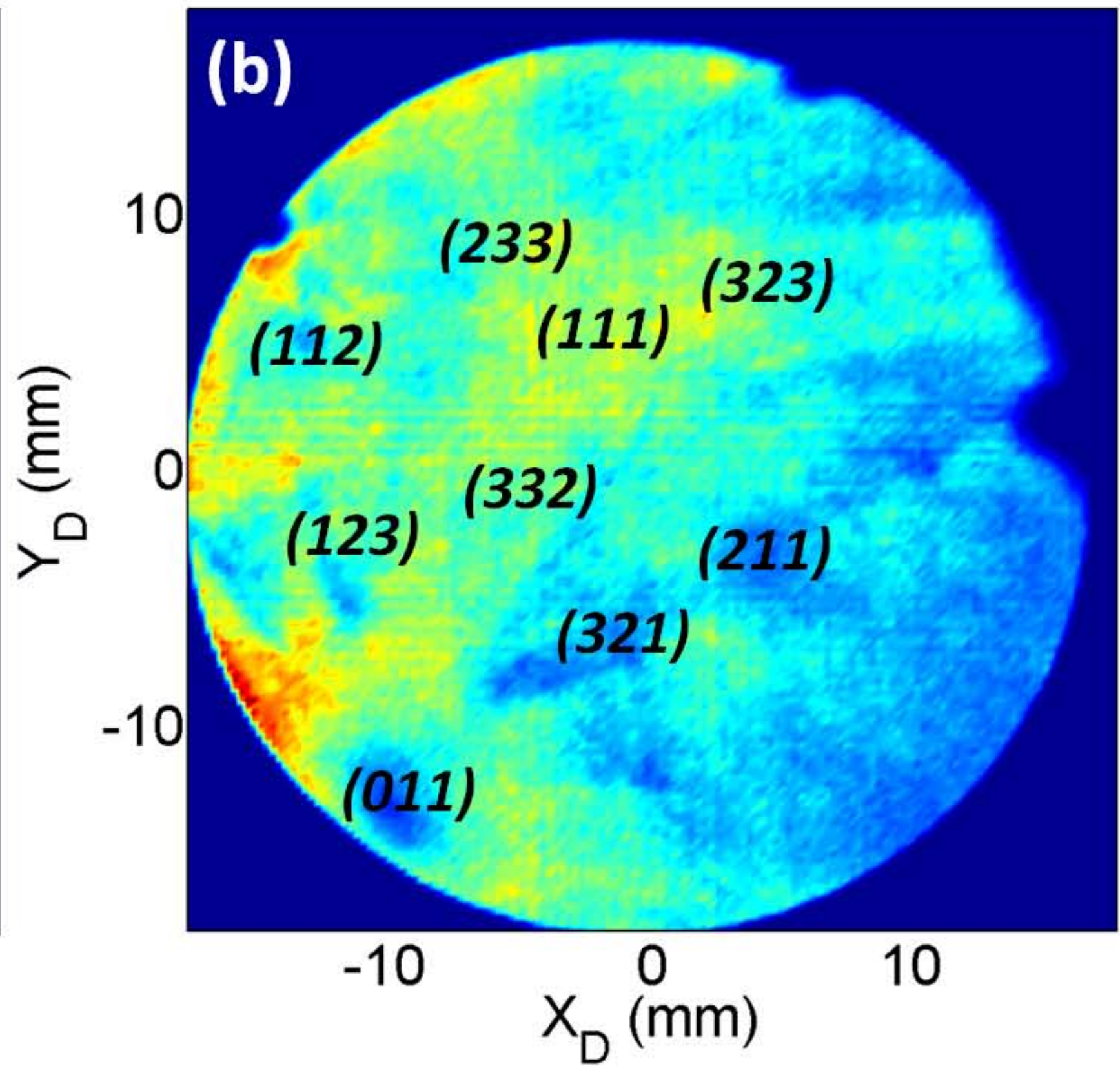

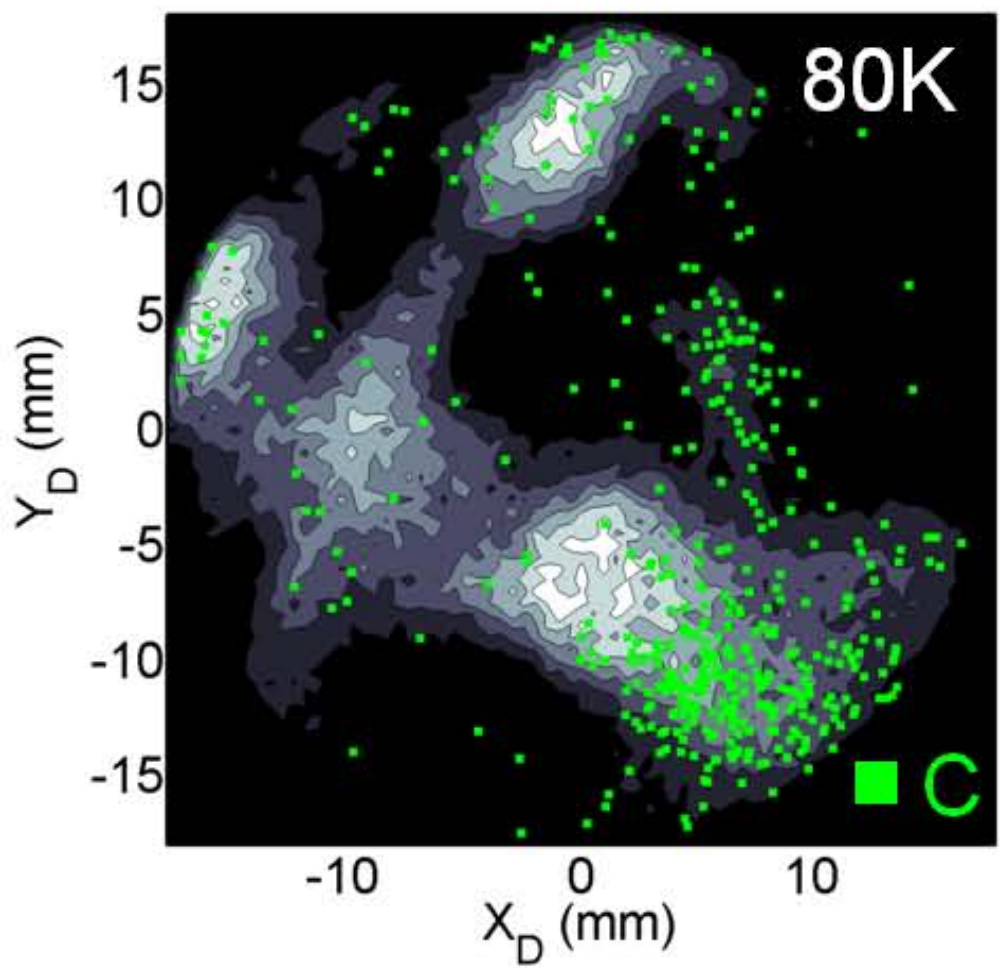
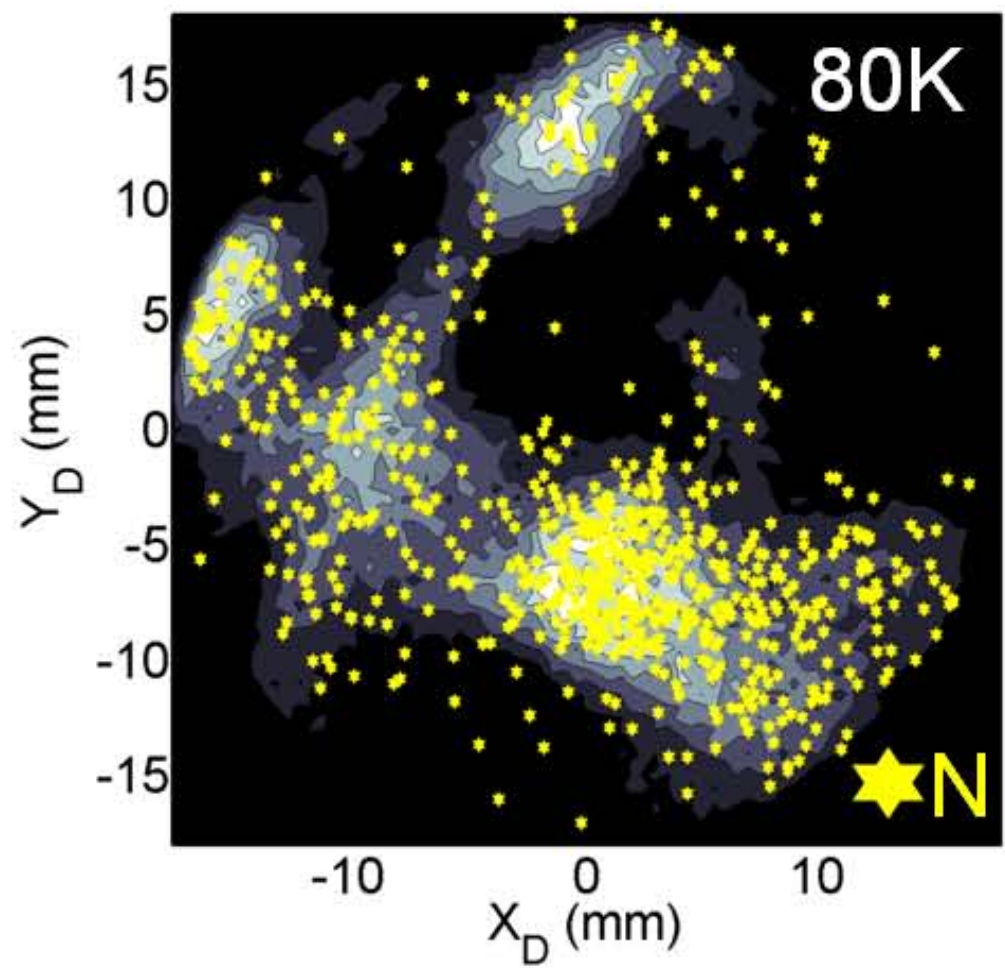
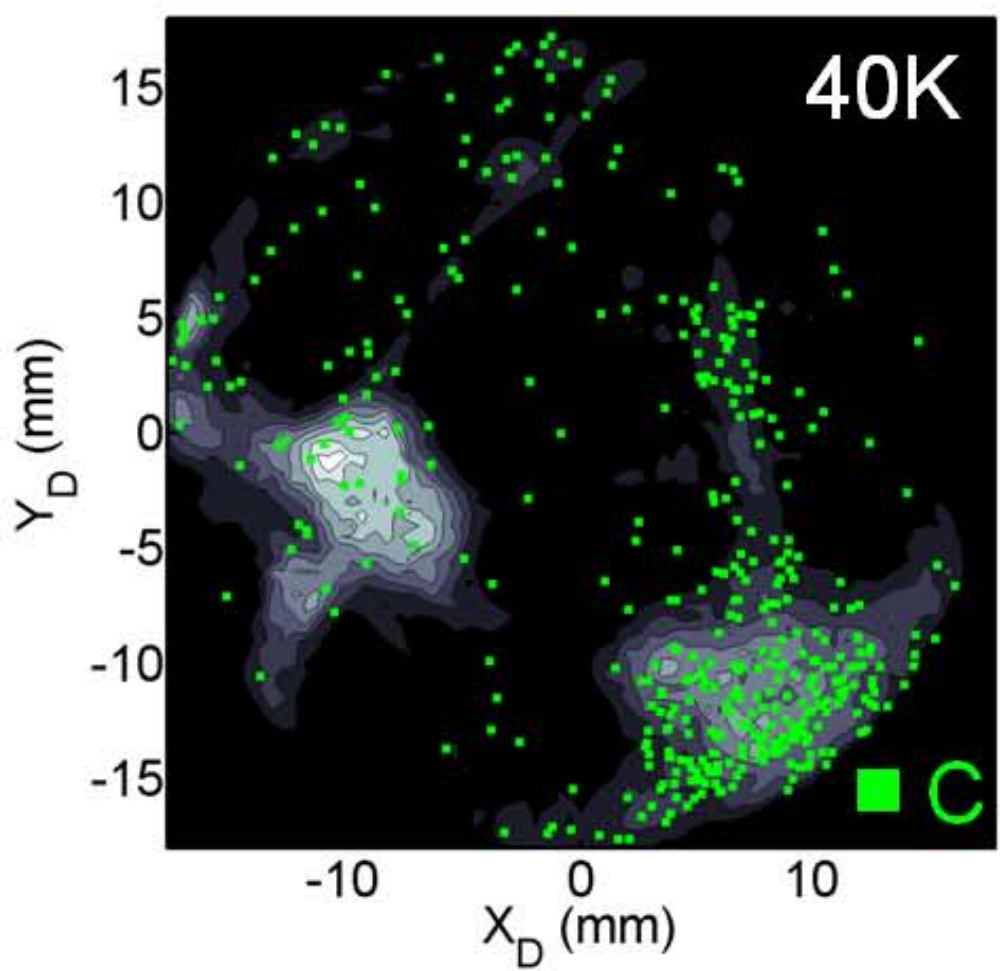
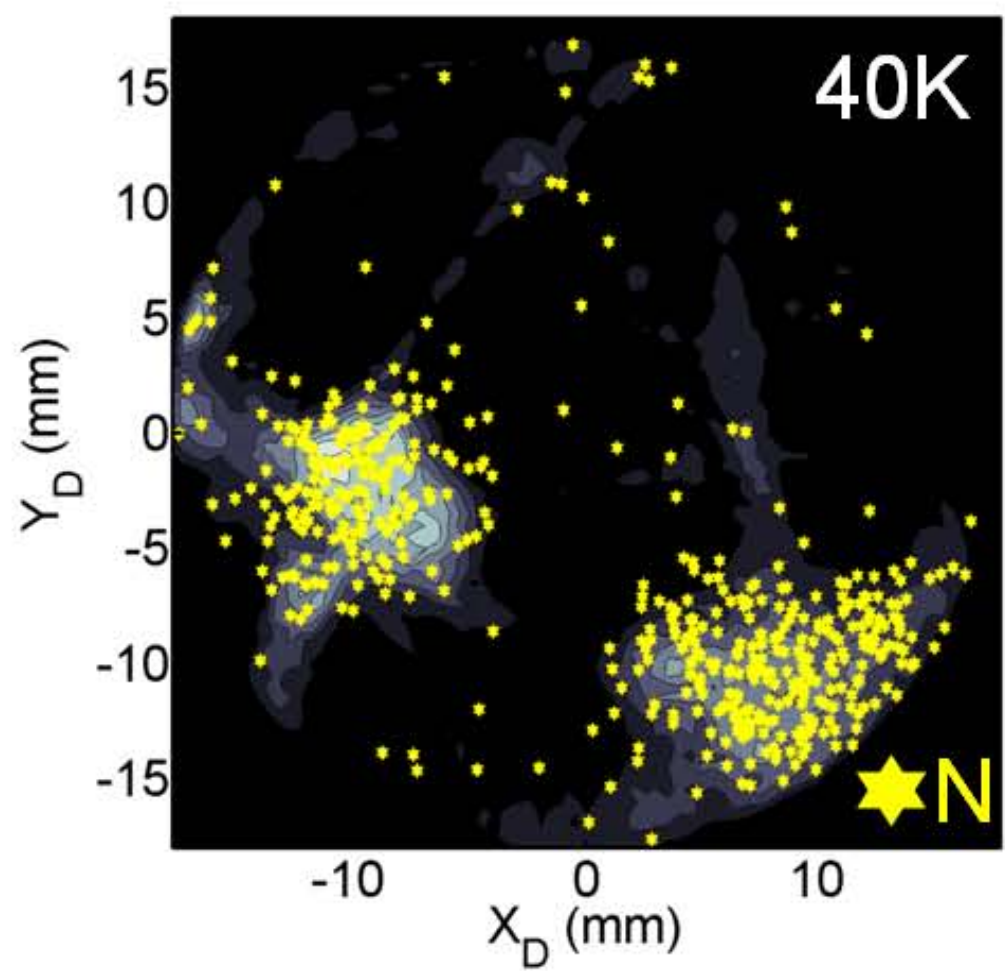
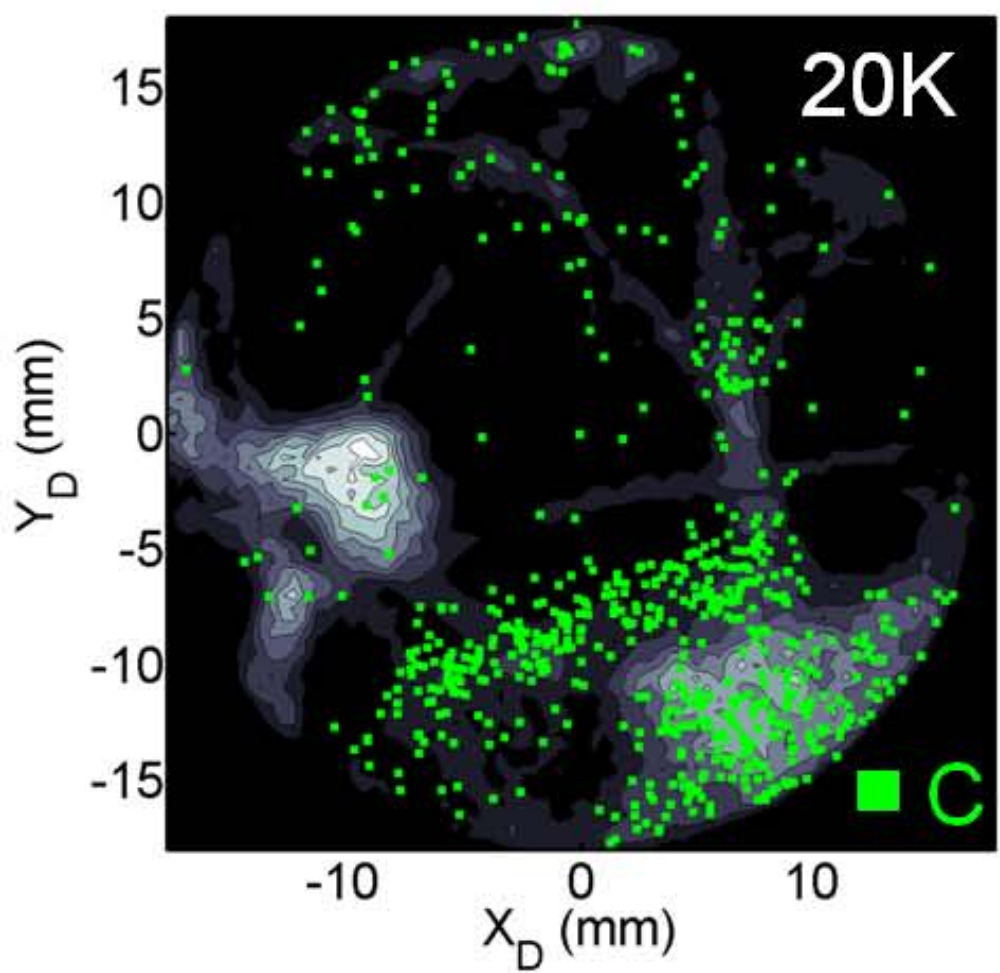
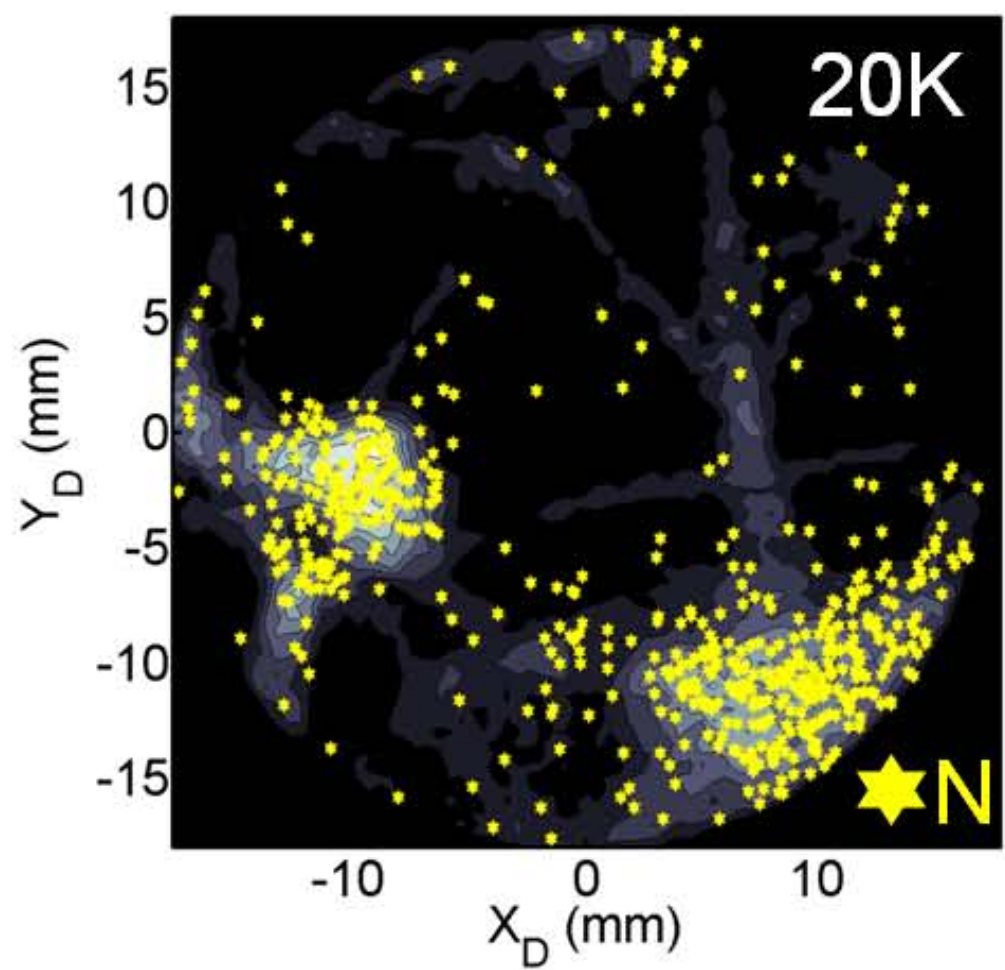

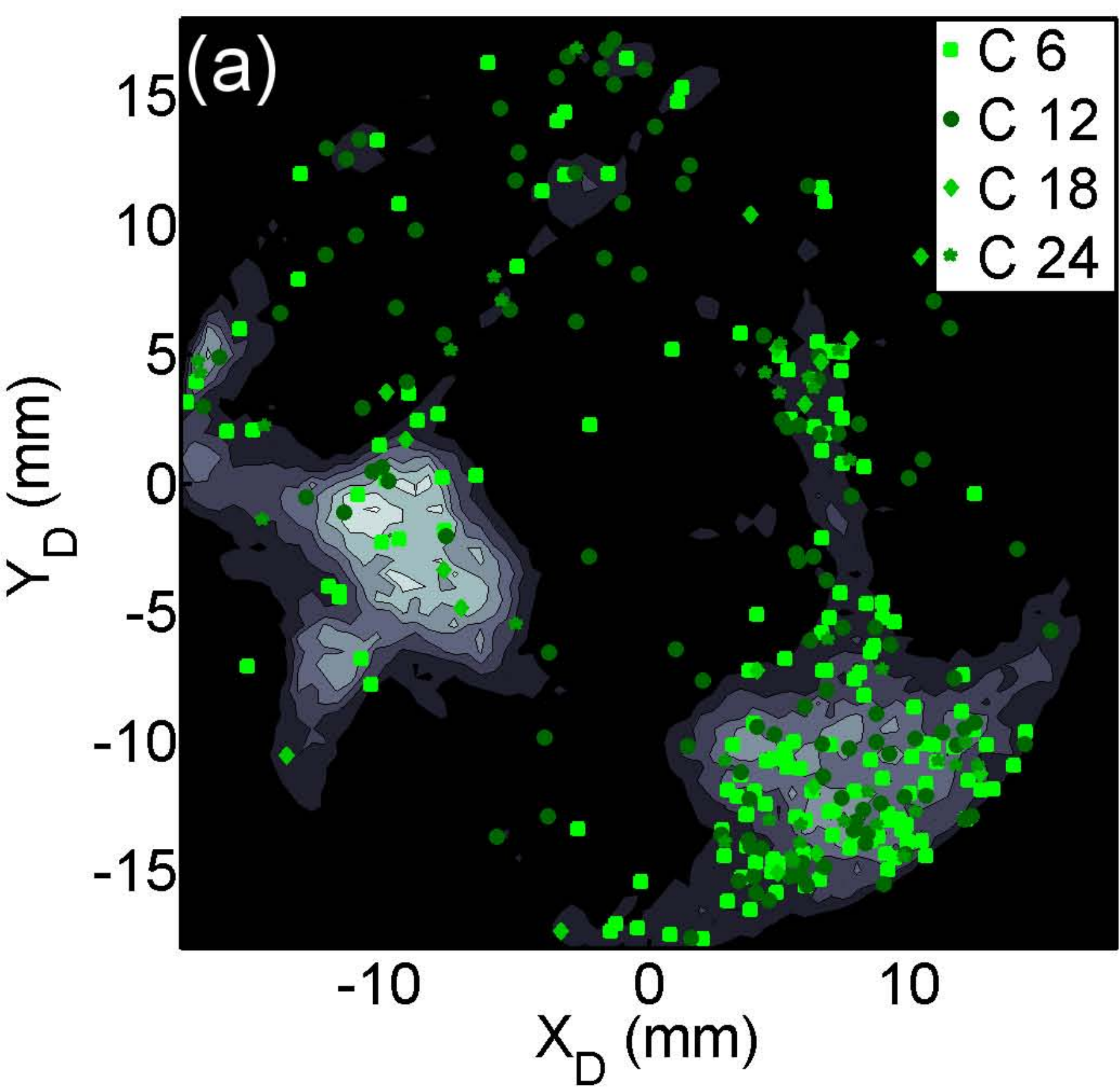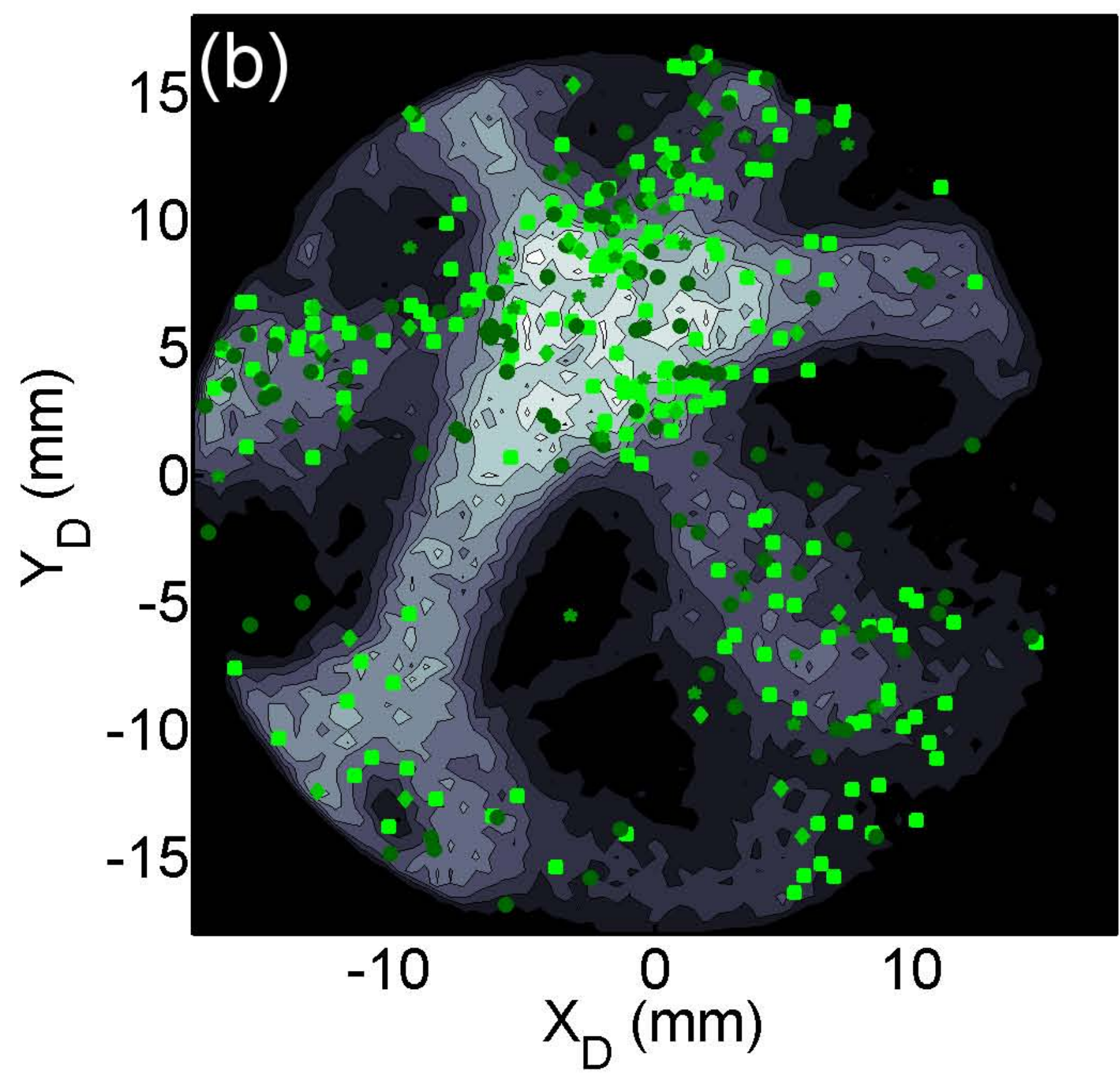

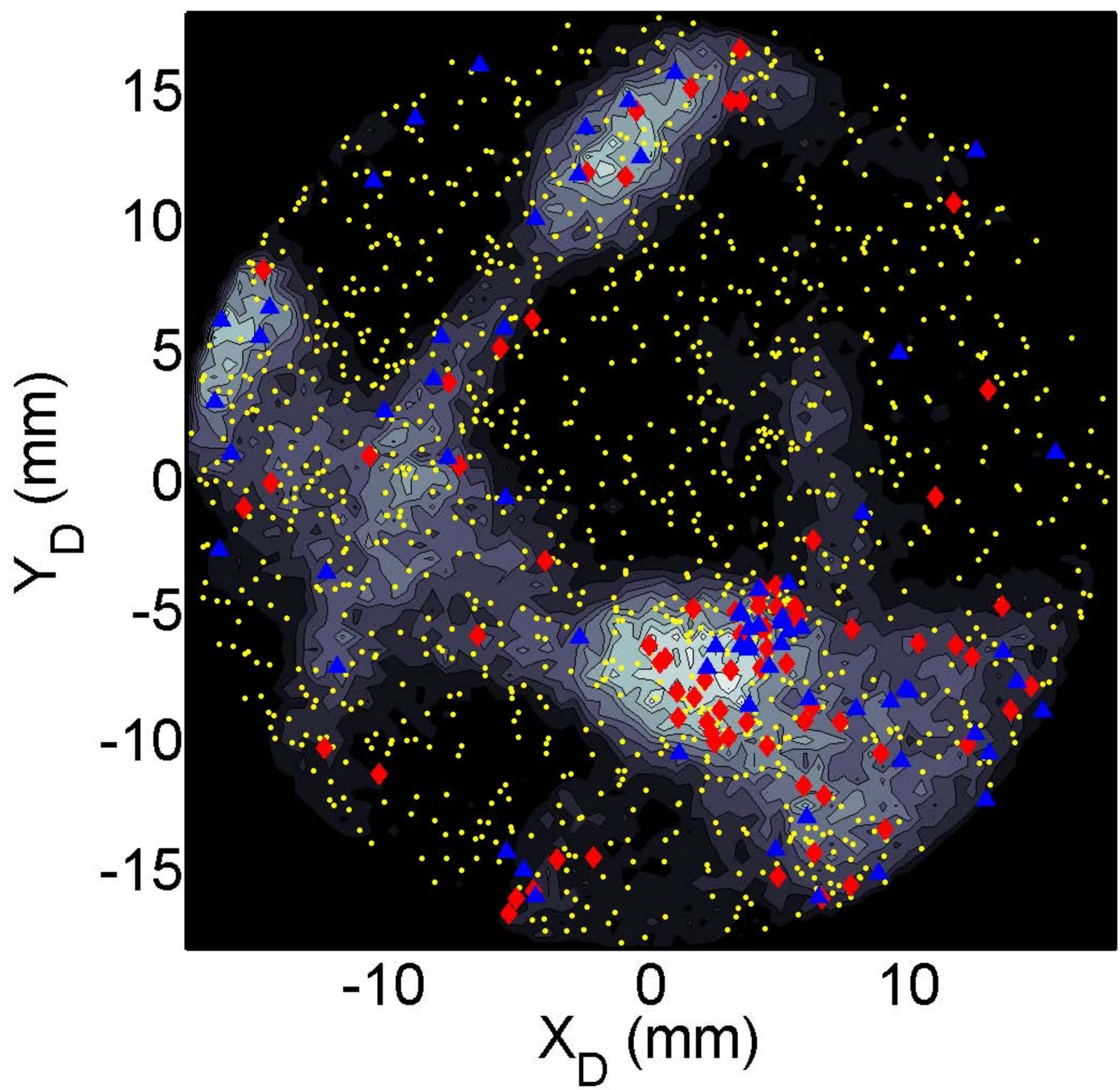

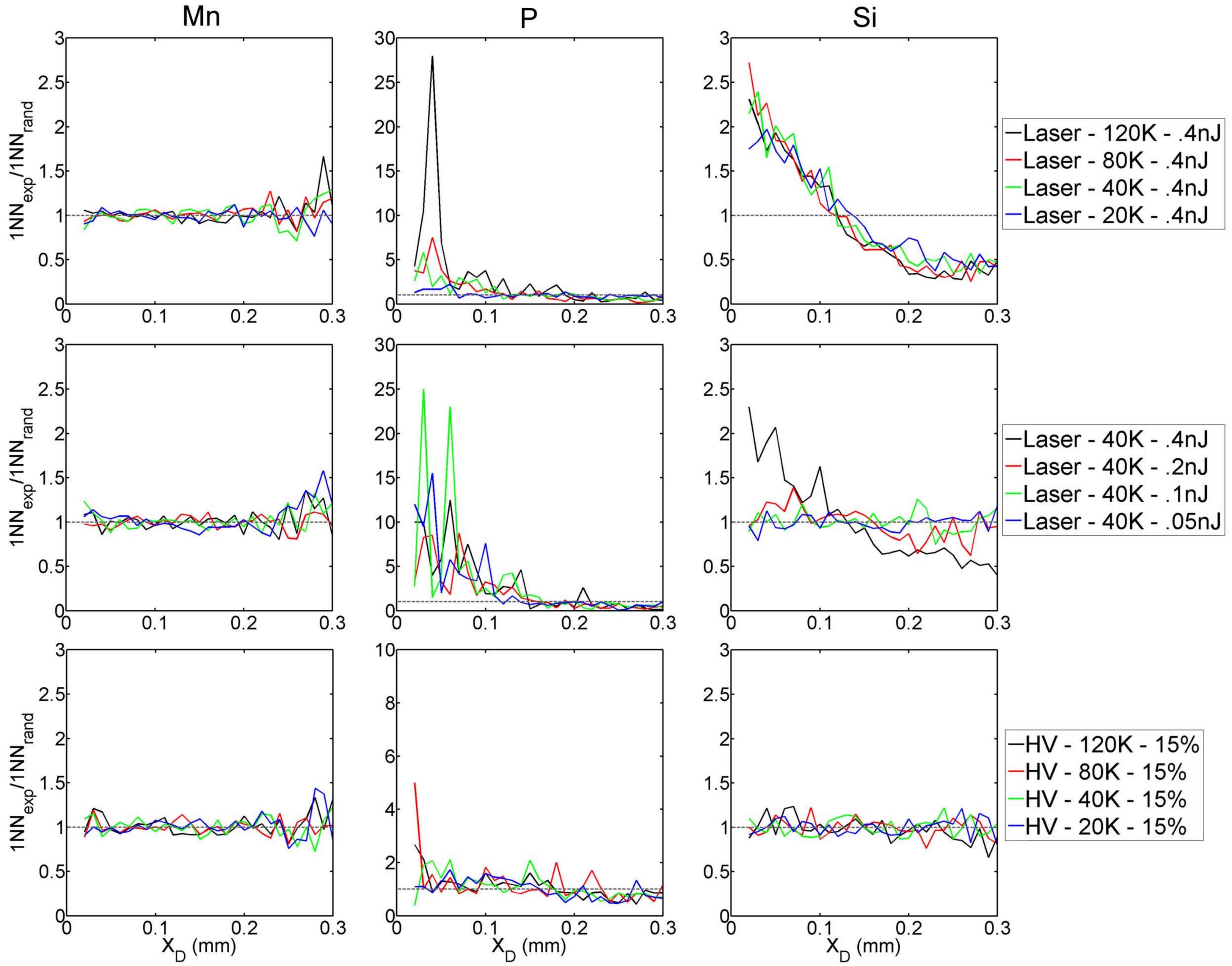

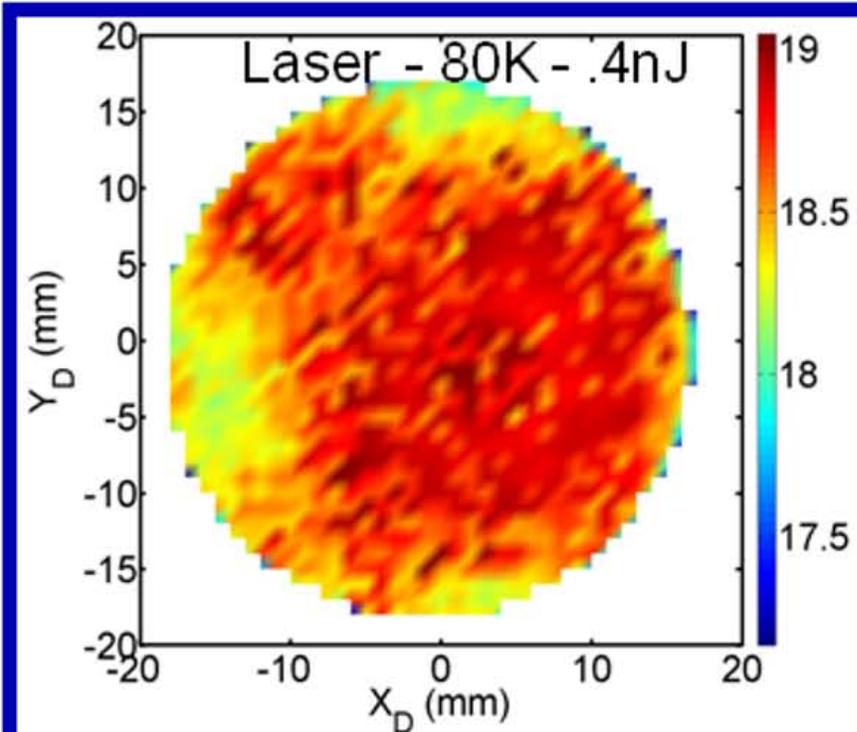
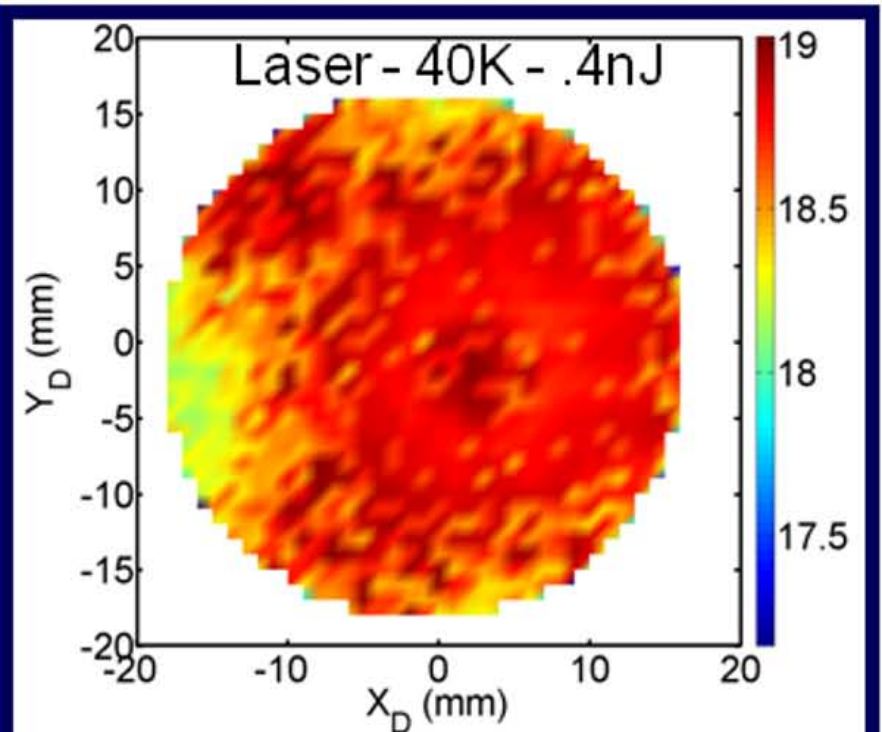
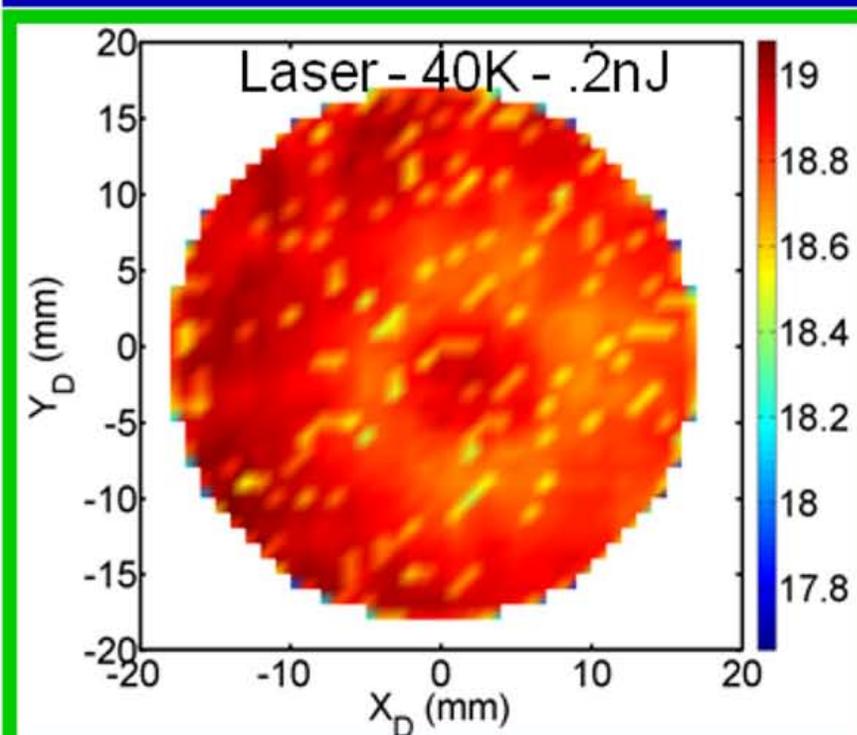
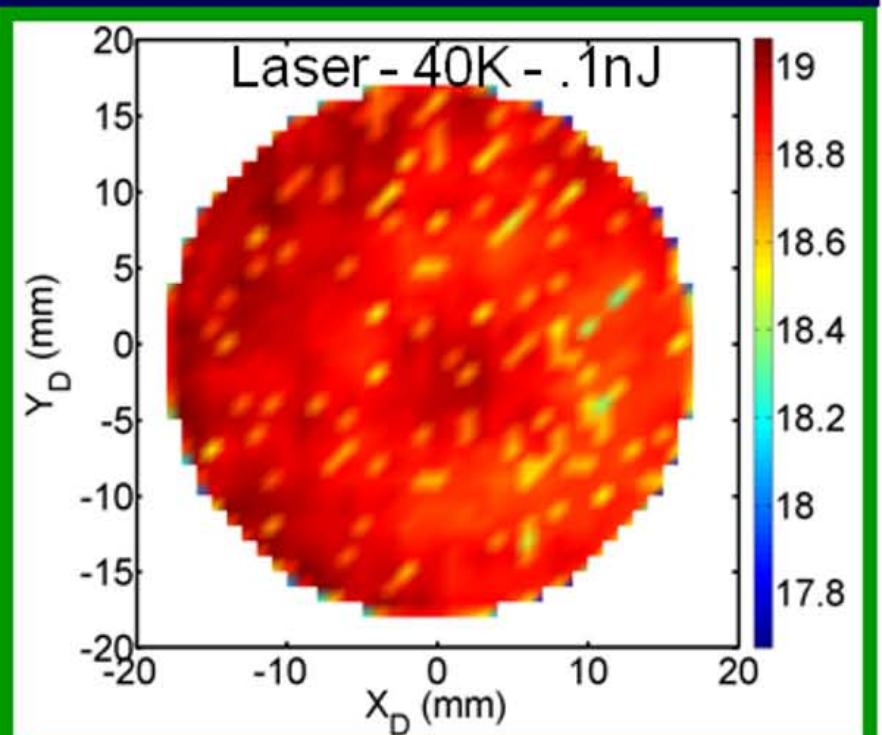
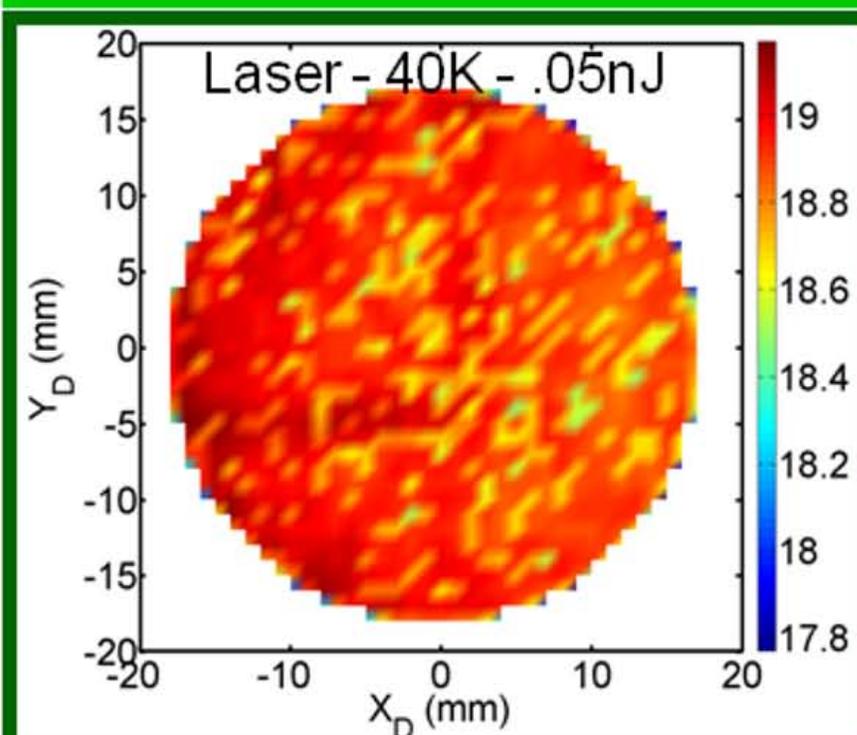
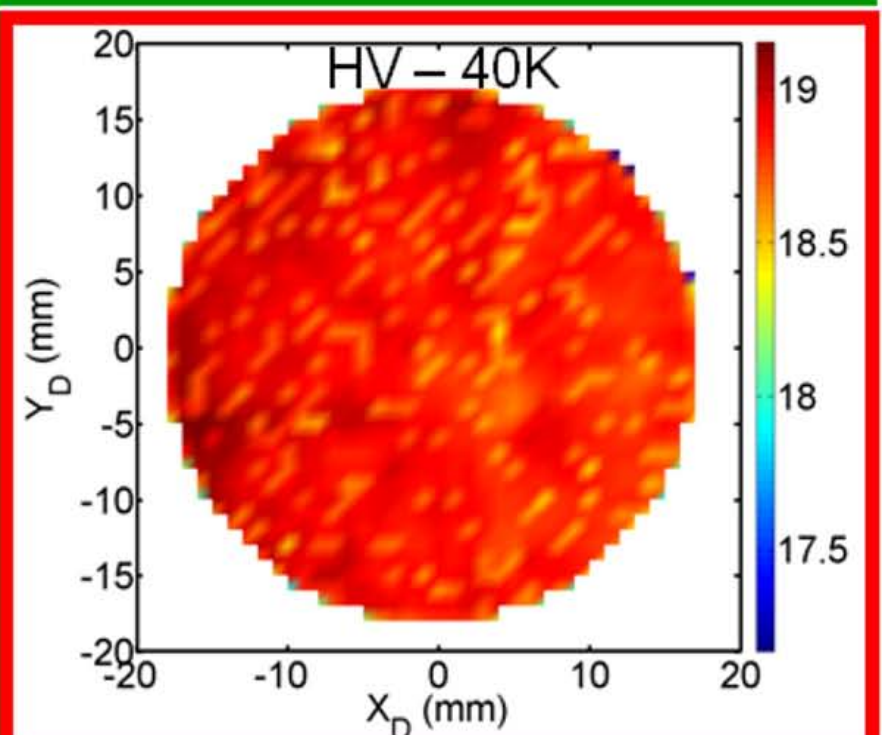

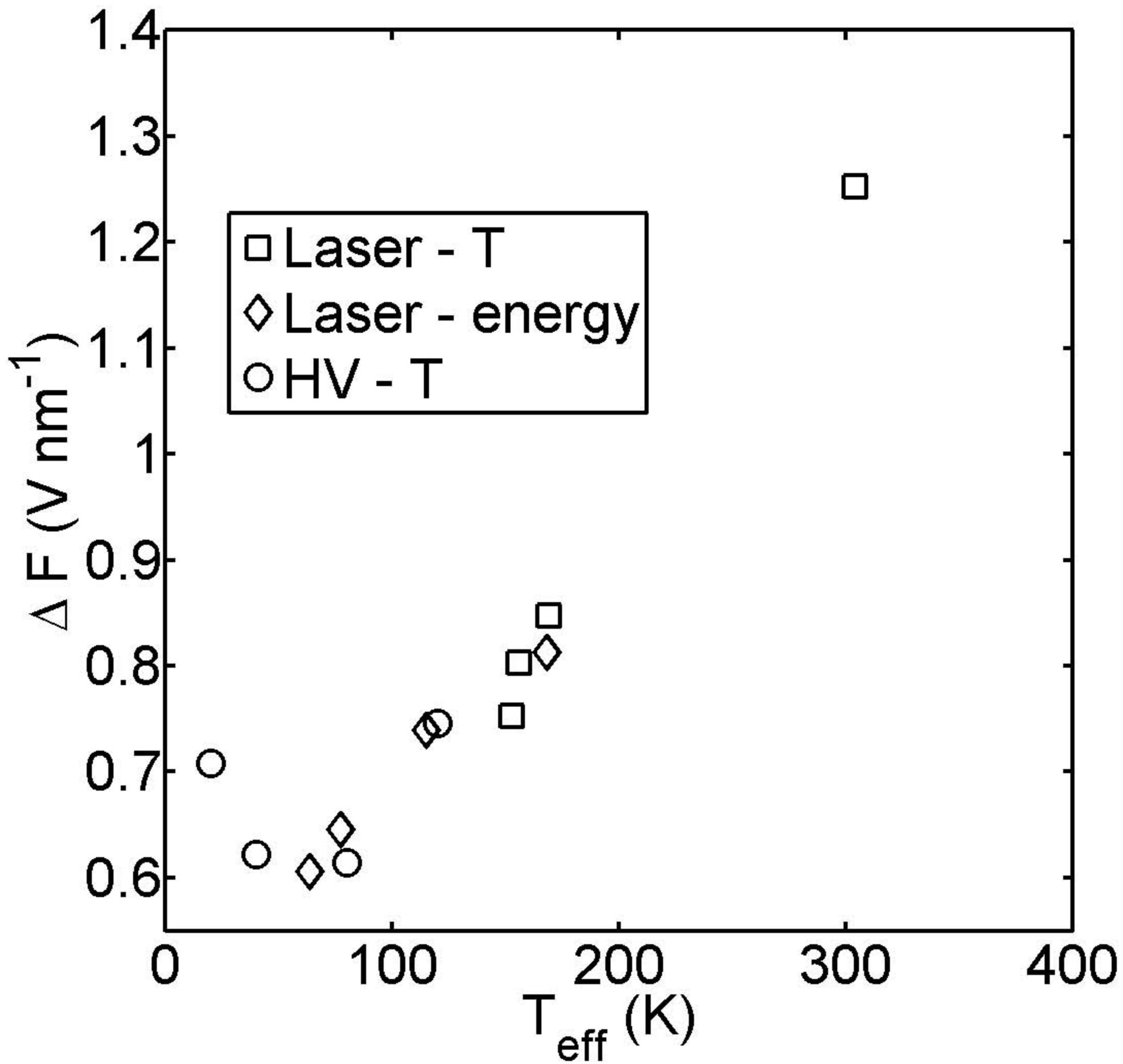

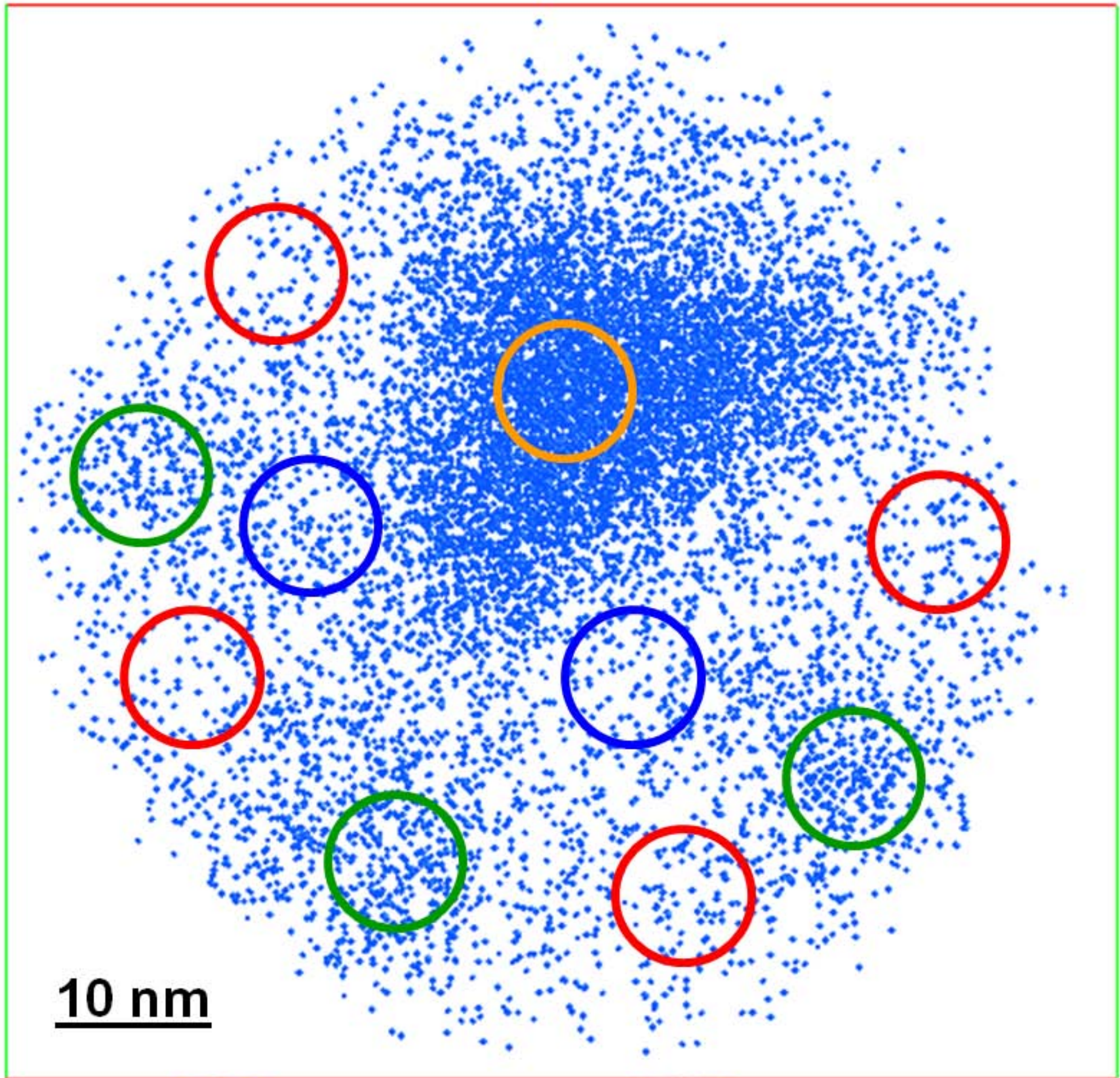

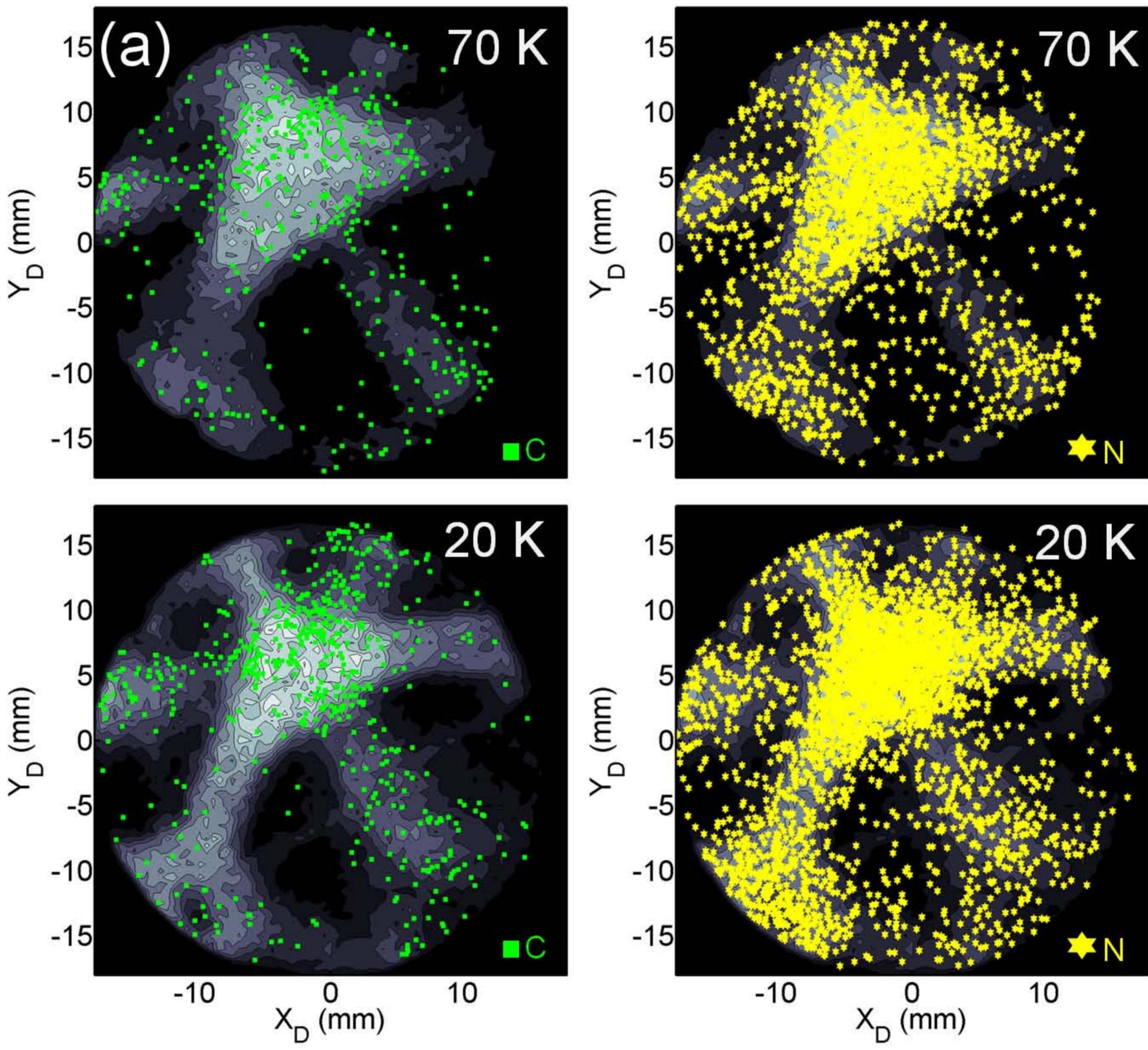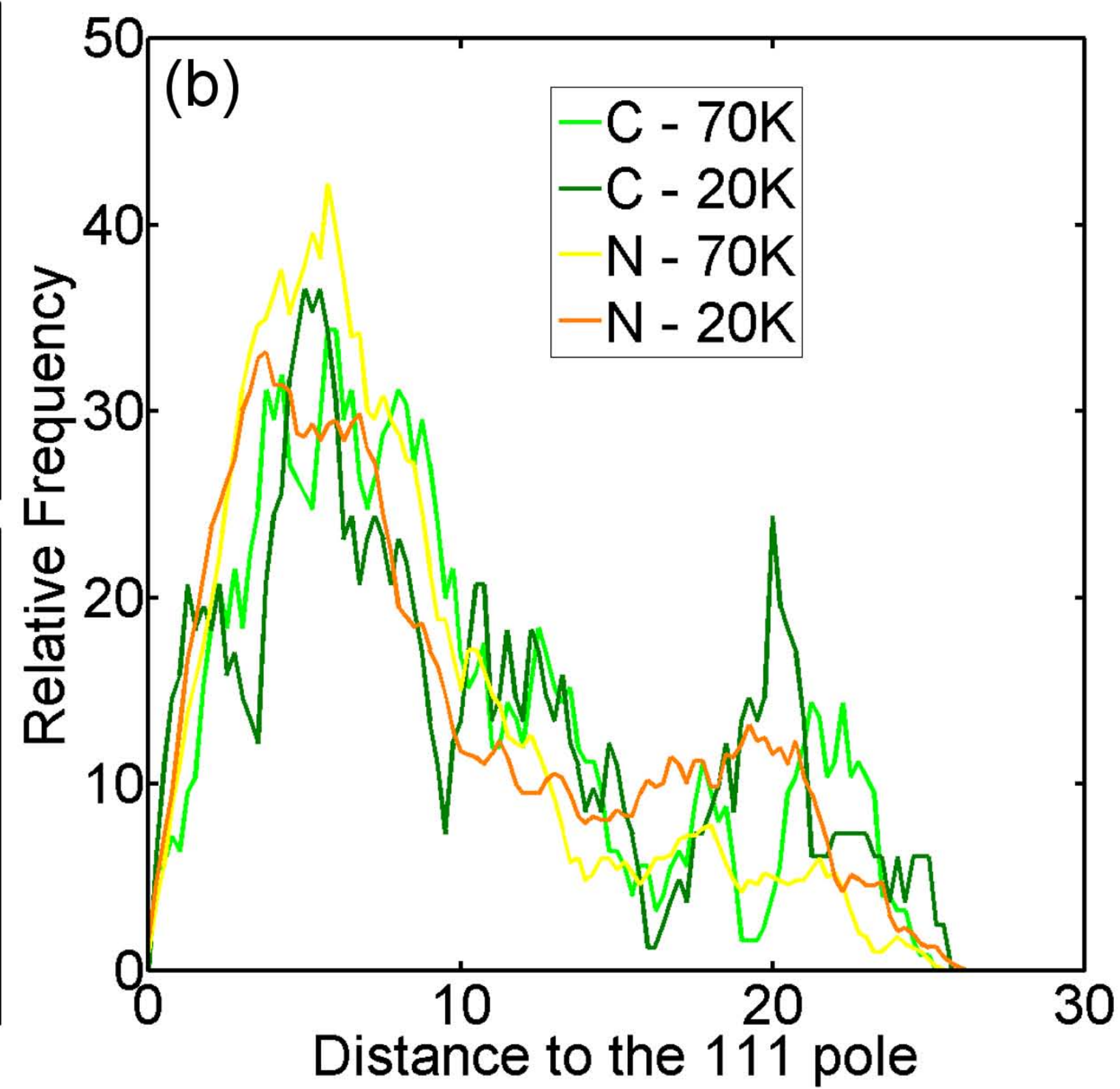